\documentstyle[twoside,epsf]{article}

\catcode`\@=11
\long\def\@makefntext#1{
\protect\noindent \hbox to 3.2pt {\hskip-.9pt
$^{{\eightrm\@thefnmark}}$\hfil}#1\hfill}       

\def\@makefnmark{\hbox to 0pt{$^{\@thefnmark}$\hss}}    

\def\ps@myheadings{\let\@mkboth\@gobbletwo
\def\@oddhead{\hbox{}
\rightmark\hfil\eightrm\thepage}
\def\@oddfoot{}\def\@evenhead{\eightrm\thepage\hfil
\leftmark\hbox{}}\def\@evenfoot{}
\def\sectionmark##1{}\def\subsectionmark##1{}}

\oddsidemargin=\evensidemargin
\newcounter{sectionc}\newcounter{subsectionc}\newcounter{subsubsectionc}
\renewcommand{\section}[1] {\vspace{12pt}\addtocounter{sectionc}{1}
\setcounter{subsectionc}{0}\setcounter{subsubsectionc}{0}\noindent
    {\tenbf\thesectionc. #1}\par\vspace{5pt}}
\renewcommand{\subsection}[1] {\vspace{12pt}\addtocounter{subsectionc}{1}
\setcounter{subsubsectionc}{0}\noindent
{\bf\thesectionc.\thesubsectionc. {\kern1pt \bfit #1}}\par\vspace{5pt}}
\renewcommand{\subsubsection}[1] {\vspace{12pt}\addtocounter{subsubsectionc}{1}
    \noindent{\tenrm\thesectionc.\thesubsectionc.\thesubsubsectionc.
    {\kern1pt \tenit #1}}\par\vspace{5pt}}

\newcounter{appendixc}
\newcounter{subappendixc}[appendixc]
\newcounter{subsubappendixc}[subappendixc]
\renewcommand{\thesubappendixc}{\Alph{appendixc}.\arabic{subappendixc}}
\renewcommand{\thesubsubappendixc}
    {\Alph{appendixc}.\arabic{subappendixc}.\arabic{subsubappendixc}}

\renewcommand{\appendix}[1] {\vspace{12pt}
        \refstepcounter{appendixc}
        \setcounter{figure}{0}
        \setcounter{table}{0}
        \setcounter{lemma}{0}
        \setcounter{theorem}{0}
        \setcounter{corollary}{0}
        \setcounter{definition}{0}
        \setcounter{equation}{0}
        \renewcommand{\thefigure}{\Alph{appendixc}.\arabic{figure}}
        \renewcommand{\thetable}{\Alph{appendixc}.\arabic{table}}
        \renewcommand{\theappendixc}{\Alph{appendixc}}
        \renewcommand{\thelemma}{\Alph{appendixc}.\arabic{lemma}}
        \renewcommand{\thetheorem}{\Alph{appendixc}.\arabic{theorem}}
        \renewcommand{\thedefinition}{\Alph{appendixc}.\arabic{definition}}
        \renewcommand{\thecorollary}{\Alph{appendixc}.\arabic{corollary}}
        \renewcommand{\theequation}{\Alph{appendixc}.\arabic{equation}}
        \noindent{\tenbf Appendix \theappendixc #1}\par\vspace{5pt}}
\newcommand{\subappendix}[1] {\vspace{12pt}
        \refstepcounter{subappendixc}
        \noindent{\bf Appendix \thesubappendixc. {\kern1pt \bfit #1}}
    \par\vspace{5pt}}
\newcommand{\subsubappendix}[1] {\vspace{12pt}
        \refstepcounter{subsubappendixc}
        \noindent{\rm Appendix \thesubsubappendixc. {\kern1pt \tenit #1}}
    \par\vspace{5pt}}

\newcommand{\textlineskip}{\baselineskip=13pt}
\newcommand{\smalllineskip}{\baselineskip=10pt}

\newcommand{\copyrightheading}[1]
    {\vspace*{-2.5cm}\smalllineskip{\flushleft
    {\footnotesize Quantum Information and Computation, Vol.~1, No.~0 (2001) 000--000 #1}\\
    {\footnotesize \copyright\kern2pt Rinton Press}\\
     }}

\def\abstracts#1#2#3{{
    \centering{\begin{minipage}{4.5in}\footnotesize\baselineskip=10pt
    \parindent=0pt #1\par
    \parindent=15pt #2\par
    \parindent=15pt #3
    \end{minipage}}\par}}

\def\keywords#1{{
    \centering{\begin{minipage}{4.5in}\footnotesize\baselineskip=10pt
    {\footnotesize\it Keywords}\/: #1
     \end{minipage}}\par}}

\renewenvironment{thebibliography}[1]
        {\frenchspacing
     \ninerm\baselineskip=11pt
         \begin{list}{\arabic{enumi}.}
        {\usecounter{enumi}\setlength{\parsep}{0pt}
     \setlength{\leftmargin 12.7pt}{\rightmargin 0pt}
         \setlength{\itemsep}{0pt} \settowidth
    {\labelwidth}{#1.}\sloppy}}{\end{list}}

\newcounter{itemlistc}
\newcounter{romanlistc}
\newcounter{alphlistc}
\newcounter{arabiclistc}

\newcommand{\fcaption}[1]{
        \refstepcounter{figure}
        \setbox\@tempboxa = \hbox{\footnotesize Fig.~\thefigure. #1}
        \ifdim \wd\@tempboxa > 5in
           {\begin{center}
        \parbox{5in}{\footnotesize\smalllineskip Fig.~\thefigure. #1}
            \end{center}}
        \else
             {\begin{center}
             {\footnotesize Fig.~\thefigure. #1}
              \end{center}}
        \fi}

\newcommand{\tcaption}[1]{
        \refstepcounter{table}
        \setbox\@tempboxa = \hbox{\footnotesize Table~\thetable. #1}
        \ifdim \wd\@tempboxa > 5in
           {\begin{center}
        \parbox{5in}{\footnotesize\smalllineskip Table~\thetable. #1}
            \end{center}}
        \else
             {\begin{center}
             {\footnotesize Table~\thetable. #1}
              \end{center}}
        \fi}

\def\pmb#1{\setbox0=\hbox{#1}
    \kern-.025em\copy0\kern-\wd0
    \kern.05em\copy0\kern-\wd0
    \kern-.025em\raise.0433em\box0}

\def\fnt#1#2{\footnotetext{\kern-.3em
    {$^{\mbox{\scriptsize #1}}$}{#2}}}

\def\fpage#1{\begingroup
\voffset=.3in
\thispagestyle{empty}\begin{table}[b]\centerline{\footnotesize #1}
    \end{table}\endgroup}

\def\runninghead#1#2{\pagestyle{myheadings}
\markboth{{\protect\footnotesize\it{\quad #1}}\hfill}
{\hfill{\protect\footnotesize\it{#2\quad}}}}
\font\tenrm=cmr10
\font\tenit=cmti10
\font\tenbf=cmbx10
\font\bfit=cmbxti10 at 10pt
\font\ninerm=cmr9

\font\eightrm=cmr8

\def\FigName{figure}%
\newbox\captionbox
\long\def\@makecaption#1#2{%
  \ifx\FigName\@captype
    \vskip\abovecaptionskip
    \setbox\tempbox\hbox{{\figurecaptionfont #1\hskip1em #2}}
    \ifdim\wd\tempbox< 28pc
    \centerline{\box\tempbox}
    \else
    {\figurecaptionfont #1\hskip1em #2\par}
\fi\else
    \setbox\tempbox\hbox{{\tablecaptionfont #1\hskip1em #2}}
    \ifdim\wd\tempbox< 28pc
    \centerline{\box\tempbox}
    \else
    {\tablecaptionfont #1\hskip1em #2\par}%
    \fi
 \vskip\belowcaptionskip
 \fi}
\InputIfFileExists{psfig.sty}
{\typeout{^^Jpsfig.sty inputed...ok}}{\typeout{^^JWarning: psfig.sty could be be found.^^J}}
\InputIfFileExists{epsfsafe.tex}
{\typeout{^^Jepsfsafe.tex inputed...ok}}
            {\typeout{^^JWarning: epsfsafe.tex could not be found.^^J}}
\InputIfFileExists{epsfig.sty}
{\typeout{^^Jepsfig.sty inputed...ok}}{\typeout{^^JWarning: epsfig.sty could not be found.^^J}}
\InputIfFileExists{epsf.sty}
{\typeout{^^Jepsf.sty inputed...ok}}{\typeout{^^JWarning: epsf.sty could not be found.^^J}}%
\def\fps@figure{tbp}
\def\ftype@figure{1}
\def\ext@figure{lof}
\def\fnum@figure{Fig.\ \thefigure}
\def\qed{\hbox{${\vcenter{\vbox{              
   \hrule height 0.4pt\hbox{\vrule width 0.4pt height 6pt
   \kern5pt\vrule width 0.4pt}\hrule height 0.4pt}}}$}}

\newcommand{\tr}{{\rm tr }}

\newcommand{\ket}[1]{\left | #1 \right \rangle}
\newcommand{\bra}[1]{\left \langle #1 \right |}

\newcommand{\proj}[1]{\ket{#1}\! \bra{#1}}

\newcommand{\bq}{\begin{quotation}\noindent}
\newcommand{\eq}{\end{quotation}}
\newcommand{\be}{\begin{equation}}
\newcommand{\ee}{\end{equation}}
\newcommand{\bea}{\begin{eqnarray}}
\newcommand{\eea}{\end{eqnarray}}
\newcommand{\bc}{\begin{center}}
\newcommand{\ec}{\end{center}}

\begin{document}

\setlength{\textheight}{8.0truein}

\runninghead{Quantumness of a Set of Quantum States}
            {C.~A. Fuchs and M. Sasaki}

\normalsize\textlineskip
\thispagestyle{empty}
\setcounter{page}{1}



\fpage{1}
\centerline{\bf Squeezing Quantum Information through a
Classical Channel:}
\vspace*{0.035truein}
\centerline{\bf Measuring the ``Quantumness'' of a Set of
Quantum States}
\vspace*{0.37truein}
\centerline{\small Christopher A. Fuchs\footnote{Address
during sabbatical year: \ Communication Networks Research Institute,
Dublin Institute of Technology, Rathmines Road, Dublin 6, Ireland.}}
\vspace*{0.015truein}
\centerline{\footnotesize\it Bell Labs, Lucent Technologies}
\baselineskip=10pt
\centerline{\footnotesize\it 600--700 Mountain Ave., Murray Hill, New
Jersey 07974, USA}
\vspace*{0.015truein}
\centerline{\footnotesize and}
\vspace*{0.015truein}
\centerline{\footnotesize\it Research Center for Quantum Communication,
Tamagawa University}
\baselineskip=10pt
\centerline{\footnotesize\it Machida, Tokyo 194-8610, Japan}
\vspace*{20pt}
\centerline{\small Masahide Sasaki}
\vspace*{0.025truein}
\centerline{\footnotesize\it Communications Research Laboratory}
\baselineskip=10pt
\centerline{\footnotesize\it Koganei, Tokyo 184-8795, Japan}
\vspace*{0.015truein}
\centerline{\footnotesize and}
\vspace*{0.015truein}
\centerline{\footnotesize\it CREST, Japan Science and Technology
Corporation}
\baselineskip=10pt
\centerline{\footnotesize\it Shibuya, Tokyo 150-0002, Japan}
\vspace*{0.225truein}

\vspace*{0.21truein}
\abstracts{In this paper we propose a general method to quantify how
``quantum'' a {\it set\/} of quantum states is.  The idea is to gauge
the quantumness of the set by the worst-case difficulty of
transmitting the states through a purely classical communication
channel. Potential applications of this notion arise in quantum
cryptography, where one might like to use an alphabet of states that
promises to be the most sensitive to quantum eavesdropping, and in
laboratory demonstrations of quantum teleportation, where it is
necessary to check that quantum entanglement has actually been used
in the protocol.}{}{}

\vspace*{10pt}
\vspace*{3pt}

\vspace*{1pt}\textlineskip

\section{Introduction}

How quantum can a single quantum state be?  Does this question make
sense?  One gets the impression that it does with even a small
perusal of the quantum-optics literature, where the coherent states
of a single mode of the electromagnetic field are often called
``classical'' states of light~\cite{Glauber63}. Despite the
nomenclature, however, we suggest that there is no robust notion of
the quantumness or classicality of a {\it single\/} quantum state.
Simply consider any two distinct coherent states $|\alpha\rangle$ and
$|\beta\rangle$ with finite $\alpha$ and $\beta$.  The inner product
of these two states is nonzero. Thus, if a single mode is prepared
secretly in one of these states, there is no automatic device that
can amplify the signal reliably into a two-mode state
$|\gamma\rangle|\gamma\rangle$, where $\gamma=\alpha,\beta$ depending
upon the input~\cite{Wootters82,Dieks82}. Nonorthogonal states cannot
be cloned~\cite{Yuen86}, and this holds whether the quantum optics
community calls such states ``classical'' or not.

Thus we feel that a robust notion of the {\it quantumness of
states\/} can only be attached to a {\it set\/} of states.  The
members of a set of states can be more or less quantum with respect
to each other, but there is no good sense in which each one alone is
intrinsically quantum or not. A set of just two nonorthogonal states
$|\psi_0\rangle$ and $|\psi_1\rangle$ provides a good
example~\cite{Fuchs00}.  The only relevant number for our purposes is
the magnitude of their inner product
$x=|\langle\psi_0|\psi_1\rangle|$. To set the stage for our later
discussion, let us once more work within the metaphor of the
no-cloning theorem.  As a warm-up, we might take the precise degree
of ``clonability''~\cite{Buzek96} as measure of the quantumness of
the two states.

There are at least two ways to pursue this~\cite{Bruss98}, but let
us focus on one for concreteness.  A cloning attempt can be
defined as any unitary operation $U$ that gives
\be
|\psi_i\rangle|0\rangle \longrightarrow |\Psi_i\rangle\;,
\label{David}
\ee
where each $|\Psi_i\rangle$ may be any (entangled) state, so long
as its partial trace over either subsystem leads to the same
density operator.  An optimal cloning attempt is one that
maximizes the fidelity between such an output and the wished-for
target state
$|\psi_i,\psi_i\rangle\equiv|\psi_i\rangle|\psi_i\rangle$. In
other words, an optimal cloning attempt is one that maximizes
\be
F_{\mbox{\scriptsize
try}}=\frac{1}{2}|\langle\Psi_0|\psi_0,\psi_0\rangle|^2 +
\frac{1}{2}|\langle\Psi_1|\psi_1,\psi_1\rangle|^2\;.
\label{Artur}
\ee
In Ref.~\cite{Bruss98}, it was shown that
\begin{eqnarray}
F_{\mbox{\scriptsize clone}}\; &=&\;
\max_U\; F_{\mbox{\scriptsize try}} \nonumber \\
&=&
\; \frac{1}{2}\Big(1+x^3+(1-x^2)\sqrt{1+x^2}\Big)\;.
\label{Christopher}
\end{eqnarray}
Viewing Eq.~(\ref{Christopher}) as a quantitative measure of the
quantumness of two states---i.e., the smaller
$F_{\mbox{\scriptsize clone}}$, the more quantum the set of
states---one finds that two states are the {\it most\/} quantum
with respect to each other when $x=1/\sqrt{3}$.

Equation~(\ref{Christopher})---though we will not adopt it as
ultimately satisfactory for our needs---exhibits some of the main
features a measure of quantumness ought to have.  In particular, by
this measure, the quantumness of two states is minimal when either
$x=0$ or $x=1$. That is, two states are the most classical with
respect to each other when they are either orthogonal or identical.
Moreover, the set is most quantum when the states are somewhere in
between, in this case when they are roughly $54.7^\circ$ apart. This
point draws perhaps the most important contrast between notions of
quantumness (as being sought here) and notions of quantum
distinguishability~\cite{Holevo73a,Helstrom76,Fuchs96a} used in
communication theory.  As an example, in the setting of classical
communication it is often important to understand the best
probability with which a signal can be guessed correctly after a
quantum measurement has been performed on the signal carrier. For the
case at hand, the measure of optimal distinguishability would then be
given by~\cite{Helstrom76}
\be
P_s = \frac{1}{2}\Big(1+\sqrt{1-x^2}\Big)\;.
\label{Chiara}
\ee
This quantity is monotone in the parameter $x$:  Two states are
the most distinguishable when $x=0$ and the least when $x=1$. No
measure of quantumness should have this character.  Instead,
quantumness should capture more of the character of how difficult
it is to make a copy of the quantum state after some of the
information about its identity has been deposited in another
system.  In the case of two very parallel quantum states, it is
quite difficult to distinguish them, but then again there is no
great need to:  A random reproduction of either of the states will
lead to almost precisely the same state as the initial one because
of their high degree of parallelism.

Nevertheless, as already alluded to, the optimal cloning fidelity in
Eq.~(\ref{Christopher}) is not completely suited to our purposes. One
reason is that the idea of cloning does not give strong guidance for
the particular construction of fidelity in Eq.~(\ref{Artur}). Under
minor modifications of this criterion, the particular $x$ for which
two states are the most quantum with respect to each other changes
drastically.  For instance, by one of the other measures considered
in Ref.~\cite{Bruss98}, two states are the most quantum with respect
to each other when $x=1/2$---that is, when they are $60^\circ$ apart.
Furthermore, with neither of these measures do we see a potentially
desirable connection between quantumness and {\it angle\/} in Hilbert
space. If there is a connection to be explored, then one might expect
that two states should be the most quantum with respect to each other
when they are $45^\circ$ apart.  This would give the pleasing slogan:
``Two states are the most classical when they are $0^\circ$ and
$90^\circ$ apart. They are most quantum when they are halfway in
between.''

For these reasons, we will stick to a metaphor more akin to
extreme eavesdropping in quantum
cryptography~\cite{Fuchs96a,Ekert94} than to quantum cloning for
defining our problem.  In particular, we consider the following
scenario. For any given set of pure states
\be
{\cal S}=\Big\{\Pi_i=|\psi_i\rangle\langle\psi_i|\Big\}\;,
\label{John}
\ee
let us act as if there is a source emitting quantum systems with
states drawn from the set according to a probability distribution
$\pi_i$. (For the present, this probability distribution should be
considered nothing more than an artifice; ultimately it will be
discarded after an optimization.) The systems are then passed on
to an eavesdropper who is required to measure them one by one and
thereafter fully discard the originals. To make sure the latter
process is enforced, we might imagine---and thus the reason for
our paper's title---that the eavesdropper really takes the form of
two people, perhaps Eve and Yves, separated by a classical
channel.  Eve may perform any quantum measurement imaginable, but
then Yves will have access to nothing beyond the classical
information obtained from that measurement for any purpose he has
in mind. Indeed, the main purpose at hand is simply to prepare a
new system in a state as ``close as possible'' to the original.

Thus the reader should be left with the imagery of a quantum state
initially living in a large river of Hilbert space, later to be
squeezed through a very small outlet, the classical channel. The
question is, how intact can the states remain in spite of this
squeezing?  To gauge the notion of intactness with respect to the
original, we will take the average fidelity~\cite{Schumacher95}
between the initial and final states. Operationally this
corresponds to Yves handing her newly prepared quantum system back
to the initial source.  The preparer there---with a record $k$ of
which state he actually prepared---simply checks the yes-no test
$\{\Pi_k,I-\Pi_k\}$ to see whether the system has kept its initial
identity or not.  The probability that it will pass the test is
the fidelity measure we are speaking of.

Considering the best measurement and resynthesis strategies Eve
and Yves can conspire to perform gets us most of the way toward a
notion of quantumness.  The final ingredient is to imagine that
the source makes this task as hard as possible for the
surreptitious team. Conceptually, we do this by adjusting the
probabilities $\pi_i$ so that the maximum average fidelity is as
small as it can be.  The resulting fidelity is what we take to be
the quantumness of the set $\cal S$.  The intuition behind this
definition is simple. It captures in a clear-cut way how difficult
the eavesdropper's task can be made for reconstructing the set of
states $\cal S$.  Moreover, it does this even disregarding the
more subtle task of quantifying how much mutual information Eve
learns about the state's identity in the process. In a way, it
captures the raw sensitivity to eavesdropping that can be imparted
to the states in $\cal S$.

The problem promoted here certainly has roots in the ``state
estimation'' scenario studied in great detail in the recent
literature~\cite{Massar95,Derka98,Latorre98,Bruss99,Vidal99,
Acin99,Sasaki01PhasEst}.
The task there is to find optimal measurements so as to maximize the
fidelity between an original input state and a reconstructed state
according to the estimation result.
The main differences are that we have relaxed the condition that the
states in $\cal S$ be associated with a uniform distribution on
Hilbert space.  Also we have added an extra optimization over the
probability distribution $\pi_i$.  As the reader will see in the
following pages, these simple generalizations and changes have the
advantage of drawing out some conceptual aspects of the
fidelity-optimization problem that seem to have been missed
previously.  Moreover, the traditional use of ``state estimation''
has been for purposes of defining a notion of distinguishability for
quantum states~\cite{Helstrom76,Chefles00}.  As explained above, this
is certainly {\it not\/} what we are trying to get at with a notion
of quantumness.\footnote{For a different take on a notion of
quantumness---this one via the issue of reversibly extracting
classical information from an ensemble in an asymptotic setting---see
Refs.~\cite{Barnum00,Hayden02}.}

The plan of the remainder of the paper is the following.  In
Section~2, we go through a more rigorous set of definitions leading
up to the quantumness of a set of states. Furthermore, we introduce
the notion of the quantumness of a Hilbert space. In Section~3, we
pause on one of the intermediate definitions---the accessible
fidelity---and develop several results on that quantity.  These
include a more convenient expression that automatically takes into
account Yves' part of the problem, and also some bounds, both on the
value of the accessible fidelity and on the measurements required for
its definition. In Section~4, we study a few examples in detail.
Among them, we show how difficult it is to derive rigorously the
quantumness of just two nonorthogonal quantum states, we study
various group covariant cases, and we demonstrate an ensemble for
which the optimal measurement for mutual information is {\it not\/}
also optimal for accessible fidelity. In Section~5, we give a small
discussion of two potential applications of the notion of
quantumness: first for designing optimally ``sensitive'' alphabets
for a new class of quantum key distribution protocols, and then for
the purpose of verifying the usage of entanglement in laboratory
demonstrations~\cite{Boschi98,Furusawa98,Braunstein00} of quantum
teleportation~\cite{Bennett93}.  We close the paper with a list of
open questions.

\section{Building an Expression for Quantumness} \label{Build}

This section is devoted to defining our notion of quantumness.  We
will approach it through a series of definitions.  Some of these
will turn out to be interesting in their own right.

To set the stage, we again imagine a source producing a sequence of
states, each drawn repeatedly from the set $\cal S$ in
Eq.~(\ref{John}) according to a probability distribution
$\pi_i$.\footnote{In this paper, we restrict the notions of
accessible fidelity and quantumness to sets $\cal S$ of pure states.
One could, of course, imagine using Uhlmann's fidelity function
\cite{Jozsa94} to make the appropriate definitions for sets of mixed
states.  The problem with this approach, however, is that then these
fidelities would lose their clean operational meaning in terms of
probabilities for an eavesdropper's going unnoticed. Thus, we leave
the interesting question of how to define mixed-state analogues of
the present quantities unaddressed here.} Such a set of states {\it
along\/} with a set of assigned probabilities, we will call an {\it
ensemble\/} $\cal P$. To be as general as possible, we place no a
priori restrictions on the number of elements in $\cal S$. Nor do we
place a restriction on $d$, the dimension of the Hilbert space ${\cal
H}_d$ where the states live. Eve, in her capacity, is imagined to
perform a single quantum measurement on each of the signals.  Most
generally, this will be some positive operator-valued measure (POVM)
${\cal E}=\{E_b\}$ \cite{Helstrom76}. The outcomes of the
measurement---indexed by the subscript $b$---will be allowed to be of
any cardinality as long as it is finite. The only (necessary)
restriction on the operators $E_b$ is that they be positive
semi-definite and combine to form a resolution of the identity,
$I=\sum_b E_b$.

Yves makes use of the information Eve obtains---some explicit
index $b$---by preparing his system in a quantum state $\sigma_b$.
Since the synthesis is based solely on classical information,
there need be no restrictions on the mapping ${\cal
M}:b\rightarrow\sigma_b$. (For instance, it need have no
connection to a completely positive linear map, etc.)  Moreover,
we even contemplate the possibility that the $\sigma_b$ are mixed
states rather than pure. This assumption corresponds to the
possibility of a randomized output strategy on the part of Yves.
The conjunction of a measurement $\cal E$ and a mapping $\cal M$
constitutes a complete protocol for the eavesdropping pair.

Now let us start to put these ingredients together.  Supposing the
source actually emits the state $\Pi_i$, and Eve obtains the
outcome $b$ for her measurement, then the fidelity Yves will
achieve in this instance is
\be
F_{b,i}= \tr(\Pi_i\sigma_b) =
\langle\psi_i|\sigma_b|\psi_i\rangle\;.
\label{Akira}
\ee
However there is no predictability of Eve's measurement outcome
above and beyond what quantum mechanics allows.  Similarly, the
most we can say about the actual state the source produces is
through the probability distribution $\pi_i$.  Therefore, the
average fidelity for this protocol will be
\bea
F_{\cal P}({\cal E},{\cal M}) &=& \sum_{b,i} p(b,i) F_{b,i}
\nonumber \\
&=& \sum_{b,i} \pi_i \tr(\Pi_i E_b) \tr(\Pi_i\sigma_b)\;,
\label{Jens}
\eea
where
\be
p(b,i) = \pi_i \tr(\Pi_i E_b)
\ee
is the joint probability for an input $i$ and an output $b$.

One convenient resting place along the road to quantumness is to
consider an optimization over Yves' strategy alone.  Thus, for a
given measurement $\cal E$ and a given ensemble $\cal P$, we
define the {\it achievable fidelity\/} with respect to the
measurement to be
\be
F_{\cal P}({\cal E})=\sup_{\cal M} F_{\cal P}({\cal E},{\cal
M})\;.
\label{Samuel}
\ee
It turns out that $F_{\cal P}({\cal E})$ can be given a
particularly pleasing analytic form.

In analogy to the quantity known as accessible
information~\cite{Holevo73a,Fuchs96a,Schumacher90} which has
turned out to be so important in the theory of quantum channel
capacities, let us define the {\it accessible fidelity\/} of the
ensemble $\cal P$ to be
\be
F_{\cal P}=\sup_{\cal E} F_{\cal P}({\cal E})\;.
\label{Jeffrey}
\ee
As a matter of practice, for all the quantities in
Eqs.~(\ref{Jens}), (\ref{Samuel}) and (\ref{Jeffrey}) we will
generally eliminate the subscript $\cal P$ whenever no confusion
can arise over the ensemble.

Finally the {\it quantumness\/} of the set $\cal S$ is defined by
\be
Q({\cal S}) = \inf_{\{\pi_i\}} F_{\cal P}\;.
\label{Eugene}
\ee
This definition has the slightly awkward property that the {\it
smaller\/} $Q({\cal S})$ is, the more quantum the set $\cal S$ is.
This, of course, could be remedied easily by subtracting the
present quantity from any constant.  However, if we wanted to
further normalize the quantumness so that, say, its value achieves
a maximum when no set of states has a higher quantumness, we would
have to make use of a (presently) unknown constant in our
definition. Thus, it seems easiest for the moment to simply remain
with Eq.~(\ref{Eugene}).

This does, however, raise an important point---indeed one of
paramount concern for the ultimate use of quantumness.  Just how
quantum can a set of states be in the most extreme case?  This
prompts the definition of the {\it quantumness of a Hilbert
space\/}:
\be
Q_d = \inf_{\cal S} Q({\cal S})\;,
\label{Charles}
\ee
where the infimum is taken over {\it all\/} sets of states living
on the Hilbert space ${\cal H}_d$.

Let us now turn to deriving some properties for these quantities.
Our first stop is the accessible fidelity.

\section{The Accessible Fidelity} \label{AccFid}

It turns out to be a rather easy matter to derive an analytic
expression for the achievable fidelity $F_{\cal P}({\cal E})$ for
any given measurement $\cal E$.  First note that Eq.~(\ref{Jens})
is linear in the variables $\sigma_b$. Thus, in any decomposition
of $\sigma_b$ into a mixture of pure states, we might as well
delete $\sigma_b$ and replace it with the most advantageous
element in the decomposition. Therefore, in the maximization in
Eq.~(\ref{Samuel}), it can never hurt to assume at the outset that
the $\sigma_b$ are pure states,
$\sigma_b=|\phi_b\rangle\langle\phi_b|$.

Rewriting Eq.~(\ref{Jens}) under this assumption, we obtain
\bea
F({\cal E},{\cal M}) &=& \sum_b\sum_i \pi_i \tr(\Pi_i
E_b)\langle\phi_b|\Pi_i|\phi_b\rangle
\nonumber\\
&=& \sum_b\sum_i \pi_i \langle\psi_i|E_b|\psi_i\rangle
\langle\phi_b|\psi_i\rangle\langle\psi_i|\phi_b\rangle
\nonumber\\
&=& \sum_b \langle\phi_b|\!\left(\sum_i
\pi_i |\psi_i\rangle\langle\psi_i|E_b|\psi_i\rangle
\langle\psi_i|\right)\!|\phi_b\rangle
\nonumber\\
&=& \sum_b \langle\phi_b|M_b|\phi_b\rangle\;,
\label{Tal}
\eea
where the Hermitian (in fact, positive semi-definite) ``mapping
operators'' $M_b$ are defined by
\bea
M_b &=& \sum_i \pi_i \Pi_i E_b\Pi_i
\nonumber\\
&=& \sum_i \pi_i \tr(\Pi_i E_b) \Pi_i\;.
\label{Eric}
\eea
(As promised, since $\cal P$ is fixed for the present
considerations, we have dropped explicit reference to it in our
notation.)  Now the pure states $|\phi_b\rangle$ are absolutely
arbitrary. Thus we can optimize each term in Eq.~(\ref{Tal})
separately.  This is done easily enough by remembering that the
largest eigenvalue $\lambda_1(A)$ of any positive semi-definite
(Hermitian) operator $A$ can be characterized by~\cite{Bhatia97}
\be
\lambda_1(A) = \sup\, \langle\alpha|A|\alpha\rangle\;,
\label{Peter}
\ee
where the supremum is taken over all normalized vectors
$|\alpha\rangle$. Therefore, the achievable fidelity reduces to
the following explicit expression:
\be
F({\cal E})=\sum_b\, \lambda_1\!\!\left(\sum_i \pi_i \tr(\Pi_i
E_b) \Pi_i\right)\;.
\label{William}
\ee

Unfortunately, this is where the easy part of the development
ends. Just as with the accessible information for an ensemble, no
explicit expression for the accessible fidelity is likely to exist
in the general case.  The best one can hope for is the
understanding of some of its general properties, a few explicit
examples, and perhaps some useful bounds.

In this regard, perhaps the first question one should ask is how
much can said about the measurements achieving equality in
Eq.~(\ref{Jeffrey}).  Are we even sure that a satisfying
measurement exists?  The answer is yes, and the reason is
essentially the same as for the existence of an optimal
measurement for the accessible information~\cite{Holevo73b}.  From
Eq.~(\ref{Peter}), it follows that
\be
\lambda_1(A+B)\le\lambda_1(A)+\lambda_1(B)\;.
\label{Carlton}
\ee
This means that $F({\cal E})$ is a convex function over the the
set of POVMs.  More formally, one can think of the set of POVMs as
equipped with a convex structure by thinking of a POVM $\cal E$ as
an infinite sequence of operators $(E_b)_{b=1}^\infty$, with only
a {\it finite\/} number of nonvanishing $E_b$.  Then a natural
convex addition operation arises by taking
\be
p{\cal E} + (1-p){\cal F}\equiv\big(p E_b + (1-p)
F_b\big)_{b=1}^\infty\;.
\label{Pranaw}
\ee
The set of POVMs constructed in this way is a compact
set~\cite{Holevo73b}.  Hence, owing the continuity
of
the function $\lambda_1(A)$, it follows that $F({\cal E})$ will
achieve its supremum on an extreme point of the set of POVMs.

Furthermore, by the reasoning of Ref.~\cite[p.~1078]{Fujiwara98}
(which is used to rederive the main result of
Ref.~\cite{Davies78}), we know that for any extreme point $\cal
E$, all the nonvanishing operators $E_b$ within it must be
linearly independent.  Thus, if the states $\Pi_i$ live on a
complex Hilbert space ${\cal H}_d$, then we can restrict the
maximization in Eq.~(\ref{Jeffrey}) to POVMs with no more than
$d^2$ outcomes. (If the Hilbert space is real, there need be no
more than $\frac{1}{2}d(d+1)$ outcomes~\cite{Sasaki99}.) Finally,
because of the subadditivity in Eq.~(\ref{Carlton}), it even holds
that these $d^2$ operators can be chosen to be rank-one---that is,
simply proportional to projectors~\cite{Fujiwara98,Davies78}. This
is a generalization of Davies' theorem for accessible information
to the present context~\cite{Davies78}.

One might wish for still a further refinement in what can be said
of optimal measurements for accessible information.  For instance,
that the number of measurement outcomes need not exceed the number
of inputs---i.e., the number of values the index $i$ can take---in
analogy to the case of quantum hypothesis
testing~\cite{Helstrom76}.  This intuition is captured in a
rhetorical way by asking, what can Eve possibly do better than
make her best guess and pass that information on to Yves?  But, if
such is the case, a proof remains to be seen.  Indeed, because of
the high degree of nonlinearity in Eq.~(\ref{Jeffrey}), there may
be counterevidence from the case of accessible information itself:
For there, there are known examples where the number of outcomes
in an optimal measurement strictly exceed the number of
inputs~\cite{Shor00}.

This brings up the question of how to draw a direct comparison
between the success probability in hypothesis testing and the
achievable fidelity for any given measurement.  The usual way of
posing the hypothesis testing problem is to assume a one-to-one
correspondence between inputs $\Pi_i$ and POVM elements $E_i$,
each element signifying the guess one should make about the
input's identity. In that way of writing the problem, the average
success probability $P_s$ takes the form
\be
P_s=\sum_i \pi_i \tr(\Pi_i E_i)\;.
\label{Kurt}
\ee
Here, however, we would be reluctant to make such a restriction on
the number of outcomes.  So, we must pose the hypothesis testing
problem in a more general way.

Suppose Eve performs a measurement $\cal E$ and observes outcome
$b$ to occur.  This new information will cause her to update here
probabilities for the various inputs according to Bayes' rule:
\be
p(i|b) = \frac{p(b,i)}{p(b)} = \frac{\pi_i\tr(\Pi_i E_b)}{\tr(\rho
E_b)}\;,
\label{Howard}
\ee
where
\be
p(b)=\tr(\rho E_b)
\label{Daniel}
\ee
is the prior probability for the outcome $b$ and
\be
\rho = \sum_i \pi_i \Pi_i
\label{Richard}
\ee
is the density operator for the ensemble $\cal P$.  The method of
maximum likelihood dictates~\cite{Renyi66} that Eve's success
probability in a decision will be maximized if she simply chooses
the value $i$ for which $p(i|b)$ is maximized.  Thus her average
success probability will be
\bea
P_s({\cal E}) &=& \sum_b p(b) \max_i\{p(i|b)\}
\nonumber\\
&=& \sum_b \max_i \{\pi_i \tr(\Pi_i E_b)\}\;.
\label{Benjamin}
\eea

Equation~(\ref{Benjamin}) can be seen to compare with the
achievable fidelity $F({\cal E})$ through a simple inequality.  In
preparation for this, note that we can also write $F({\cal E})$ in
the form
\be
F({\cal E})=\sum_b p(b) \lambda_1(\rho_b)\;,
\label{Asher}
\ee
where
\be
\rho_b = \sum_i p(i|b) \Pi_i \;.
\label{Hoi-Kwong}
\ee
Now, suppose that $j$ is the index that maximizes $p(i|b)$.  Then
\bea
\lambda_1(\rho_b) &=&
\max_{|\phi\rangle}\langle\phi|\rho_b|\phi\rangle
\nonumber\\
&\ge& \langle\psi_j|\rho_b|\psi_j\rangle
\nonumber\\
&=& \sum_i p(i|b) |\langle\psi_j|\psi_i\rangle|^2
\nonumber\\
&=& p(j|b) + \sum_{i\ne j} p(i|b) |\langle\psi_j|\psi_i\rangle|^2
\nonumber\\
&\ge& \max_i \{p(i|b)\}\;.
\label{Howard2}
\eea
Therefore, we find the simple inequality we were seeking:
\be
P_s({\cal E})\le F({\cal E})\;.
\label{Jerome}
\ee
Of course, this inequality is not tight at all:  For instance, for
$\cal P$ describing a uniform distribution of states on a qubit,
$P_s\rightarrow0$, while $F$ can be as large as $2/3$
\cite{Chefles00}. However, it does show the extent to which Yves
stands a chance of recovering from Eve's measurement errors by the
generation of a new quantum state.

Tighter bounds, both upper and lower, on Eq.~(\ref{Jeffrey}) would
be very useful.  For the present, though, we have little to report
in that regard.  An obvious lower bound comes directly from the
convexity of the $\lambda_1(A)$ function.  Note in particular that
\be
\rho = \sum_b p(b) \rho_b\;.
\label{Ruediger}
\ee
Therefore
\be
F({\cal E})\ge \lambda_1(\rho)\;.
\label{Joseph}
\ee
This inequality is generally tighter than our previous in that it
never falls below $1/d$, and moreover there is a measurement $\cal
E$ that achieves it---namely choosing $\cal E$ to be the one
element set consisting of the identity operator $I$.

A more interesting lower bound to the accessible fidelity comes
about by considering the behavior of $F({\cal E})$ with respect to
the so-called ``square-root measurement'' or ``pretty good
measurement'' \cite{Holevo78,Hausladen94}.  This POVM is
constructed directly from the ensemble decomposition of $\rho$ in
Eq.~(\ref{Richard}) by multiplying it from the left and right by
$\rho^{-1/2}$.  With this we obtain a natural decomposition of the
identity:
\be
I=\sum_i \pi_i \rho^{-1/2} \Pi_i \rho^{-1/2}\;.
\label{Steven}
\ee
(To ensure that $\rho^{-1/2}$ is well defined, we restrict all
operators to the support of $\rho$; clearly there is no loss in
generality for our problem in doing this.)  Inserting this expression
into Eq.~(\ref{William}), we obtain the lower bound
\be
F_{\cal P}\ge F_{\rm SRM}=\sum_i \lambda_1\!\!\left(\sum_j \pi_i\pi_j
\Pi_j \rho^{-1/2}\Pi_i\rho^{-1/2}\Pi_j\right).
\label{Jeroen}
\ee
It seems that if one is going to find good bounds on the quantumness
of a set of states and the quantumness of a Hilbert space, the study
of this expression may be a good place to start.  However, for most
particular ensembles this bound is unlikely to be tight as a bound on
the accessible fidelity itself.

\section{Some Examples} \label{Examples}

\subsection{The Quantumness of Two Nonorthogonal States}

Demonstrating the quantumness of even simple sets of quantum
states can be surprisingly difficult.  A case in point is the
minimal set $\cal S$, consisting of just two nonorthogonal quantum
states $\ket{\psi_0}$ and $\ket{\psi_1}$.  Supposing equal
probabilities for the states, one would think intuitively that the
best Eve could do in the imaginary scenario motivating $Q$ is to
measure the observable that optimizes her error about which state
is in front of her---i.e., the Helstrom measurement leading to
Eq.~(\ref{Chiara})---and then use the procedure between
Eqs.~(\ref{Tal}) and (\ref{William}) to generate the new states
$\ket{\phi_b}$ for her ultimate output.  And that indeed is the
case.  But the trail from this intuition to a formal proof is not
completely straightforward.

We start by finding the accessible fidelity $F$ for a general
binary pure-state ensemble
\be
{\cal P}=\Big\{ \ket{\psi_0},\ket{\psi_1} ; \pi_0, \pi_1\Big\}\;.
\ee
For ease, let us define an orthonormal basis $\{ \ket{+}, \ket{-}
\}$ so that
\bea
\ket{\psi_0}&=&\cos{\theta\over2}\ket{+}
                        -\sin{\theta\over2}\ket{-},\\
\ket{\psi_1}&=&\cos{\theta\over2}\ket{+}
                        +\sin{\theta\over2}\ket{-},
\eea
where $0\le\theta\le\pi$. Clearly the only thing that should
matter about these states is the parameters $x$ and $\theta$
defined by $\langle\psi_0|\psi_1\rangle=\cos\theta=x$. According
to Eqs.~(\ref{Jeffrey}) and (\ref{William}), the achievable
fidelity for $\cal P$ with respect to a POVM $\cal E$ is given by
\be
F({\cal E})=\sum_b\lambda_1(M_b),
\label{Sasaki_Facc_def}
\ee
where
\be
M_b=\pi_0 \tr(E_b\Pi_0)\Pi_0+\pi_1 \tr(E_b\Pi_1)\Pi_1
\ee
where $\Pi_i=\proj{\psi_i}$. After a small calculation, one can
show that the two eigenvalues for each $M_b$ are given by
\be
\lambda_{\pm}(M_b)= {1\over2} \left[ \tr(E_b\rho)\pm
\sqrt{\left[\tr(E_b\rho)\right]^2\cos^2\theta
        +\left[\tr(E_b\Delta)\right]^2\sin^2\theta}\,
\right] ,
\ee
where
\be
\rho = \pi_0 \Pi_0+\pi_1 \Pi_1\qquad \mbox{and}\qquad \Delta =
\pi_0 \Pi_0-\pi_1 \Pi_1\;.
\ee
Therefore the accessible fidelity, Eq.~(\ref{Jeffrey}), reads
\be
F=\frac{1}{2}+\frac{1}{2}G\;,
\ee
where
\be
G=\max_{\cal E} \sum_b
\sqrt{\left[\tr(E_b\rho)\right]^2\cos^2\theta
        +\left[\tr(E_b\Delta)\right]^2\sin^2\theta}\;.
\label{Sasaki_Facc_exp}
\ee

The greatest obstacle for finding a general expression for $G$ is
in the nonlinearity of the function on the right-hand side of
Eq.~(\ref{Sasaki_Facc_exp}).  In principle, by the generalization
of Davies' theorem presented in Section~3, we must search over all
3-outcomed POVMs (with rank-1 elements) to find its maximal value.
However, this is an almost impossible technique to carry out
directly. We therefore resort to deriving the maximization by
first demonstrating an upper bound on $G$ and then returning to
show that equality can be achieved in each step of its derivation
with a rather simple 2-outcomed POVM---indeed, a POVM that is none
other than Helstrom's measurement.

The first step in the process is to rewrite
Eq.~(\ref{Sasaki_Facc_exp}) as
\be
G=\max_{\cal E} \sum_b \tr(E_b\rho) \sqrt{\cos^2\theta
        +\left[
          {{\tr(E_b\Delta)}\over{\tr(E_b\rho)}}
          \right]^2 \sin^2\theta}\;.
\label{Sasaki_Facc_exp2}
\ee
Notice that $\tr(\rho E_b)$ is a probability distribution over the
index $b$ and that the square root is a monotonically increasing
concave function.  Therefore, we can use Jensen's inequality to
upper bound $G$ by
\bea
G&\le&\max_{\cal E} \sqrt{ \sum_b \Biggl[
           \tr(E_b\rho)\cos^2\theta
        +{ {\left[\tr(E_b\Delta)\right]^2}
            \over{\tr(E_b\rho)}}
           \sin^2\theta
                       \Biggr] }
\label{Sasaki_Facc_bound1A}
\\
&=&
\sqrt{ \cos^2\theta
        +A\,\sin^2\theta}\;,
\label{Sasaki_Facc_bound1B}
\eea
where
\be
A=\max_{\cal E}\sum_b { {\left[\tr(E_b\Delta)\right]^2} \over
{\tr(E_b\rho)} }\;.
\ee
Equality in Eq.~(\ref{Sasaki_Facc_bound1A}) is achieved if and
only if
\be
\left[
          {{\tr(E_b\Delta)}\over{\tr(E_b\rho)}}
          \right]^2
={\rm constant}\quad\forall\; b.
\label{equality_condition1}
\ee

To find $A$, we apply a technique used in
\cite{BraunsteinCaves94,FuchsCaves94}. This is based on putting an
upper bound on $A$ in such a way that the $\tr(E_b\rho)$ term in
the denominator is cancelled and only an expression linear in
$E_b$ is left behind. For this purpose, we introduce an operator
$X$ defined implicitly as a solution to the linear equation
\be
\rho X + X \rho = 2\Delta\;.
\ee
In terms of an orthonormal basis $\{ \ket{\omega_+},\ket{\omega_-}
\}$ that diagonalizes $\rho$, $X$ can be represented by
\be
 X =
2\sum_{i,j=+,-} { {\langle\omega_i\vert\Delta\vert\omega_j\rangle}
\over {\omega_i+\omega_j} }
\vert\omega_i\rangle\langle\omega_j\vert
\label{super-op_rep}
\ee
so long as $\rho$ is invertible. The property we need is
\bea
\vert\tr(\rho E_b  X )\vert
&\ge&
{\rm Re}[ \tr(\rho E_b  X ) ]
\nonumber\\
&=&
{1\over2} \left[ \tr(\rho E_b  X ) + \tr(\rho E_b
 X)^\ast \right]
\nonumber\\
&=&\tr(E_b\Delta)\;.
\eea
By using this, the quantity $A$ is bounded as
\bea
A
&\le&
\max_{\cal E}\sum_b { {\vert\tr(\rho E_b
 X )\vert^2}
\over {\tr(E_b\rho)}}
\label{Sasaki_Facc_bound2}
\\
&=&
\max_{\cal E}\sum_b { {\left|\tr\Bigl(\rho^{1/2} E_b^{1/2}\cdot
E_b^{1/2}
 X \rho^{1/2}\Bigr)\right|^2}
\over {\tr(E_b\rho)}}
\nonumber\\
&\le&
\max_{\cal E}\sum_b \tr\Bigl( \rho^{1/2} X E_b^{1/2}\cdot
E_b^{1/2} X \rho^{1/2} \Bigr)
\label{Sasaki_Facc_bound3}
\\
&=&\tr(X\rho X)
\label{Sasaki_Facc_bound4}
\\
&=&\tr(\Delta X)\;.
\label{Ned}
\eea
Step (\ref{Sasaki_Facc_bound3}) relies on the Schwarz inequality
\be
\vert\tr( A^\dagger  B)\vert^2\le \tr( A^\dagger  A)\tr( B^\dagger
B)\;,
\ee
and the final step (\ref{Sasaki_Facc_bound4}) follows from the
fact that the operators in $\cal E$ form a resolution of the
identity.

Equality between Eqs.~(\ref{Sasaki_Facc_bound2}) and (\ref{Ned})
is achieved when the following two conditions are satisfied. From
step (\ref{Sasaki_Facc_bound2}),
\be
{\rm Im}[\tr(\rho E_b  X)]=0 \quad\forall\; b,
\label{equality_condition2}
\ee
and from step (\ref{Sasaki_Facc_bound3}),
\be
E_b^{1/2} X \rho^{1/2} =\mu_b E_b^{1/2} \rho^{1/2} \quad\forall\;
b,
\label{equality_condition3}
\ee
with some constants $\mu_b$. The second condition
Eq.~(\ref{equality_condition3}) can be met easily by choosing the
POVM $\cal E$ to consist of rank-one projectors onto a set of
eigenvectors for the Hermitian operator $ X $ and choosing the
constants $\mu_b$ to be the associated eigenvalues. The first
condition Eq.~(\ref{equality_condition2}) then follows since
\bea
\tr(\rho E_b  X)
&=&
\tr\Big(\rho E_b^{1/2} E_b^{1/2} X \Big)
\nonumber\\
&=&\mu_b\,\tr(\rho E_b),
\eea
which is clearly a real number.

Now, all that remains is to show that the POVM ${\cal E}=\{E_b\}$
constructed in this way also satisfies
Eq.~(\ref{equality_condition1}) and hence is an optimal solution
for the accessible fidelity.

Let us define (without loss of generality) a parameter
\be
P\equiv\pi_2-\pi_1\ge0, \quad (0\le\pi_1\le1/2)\;.
\ee
Then, the spectral decomposition of $\rho$ is given by
\be
\rho=\sum_{i=+,-}\omega_i\proj{\omega_i}, \quad
\ket{\omega_i}=U\ket{i},
\ee
where
\be
\omega_{\pm}={1\over2} \left( 1\pm\sqrt{\cos^2\theta + P^2
\sin^2\theta} \right),
\ee
and
\be
U= \pmatrix{ \cos\gamma & -\sin\gamma\cr \sin\gamma
& \cos\gamma }
\ee
with
$$
\cos^2\gamma = {{ \sqrt{ \cos^2\theta + P^2 \sin^2\theta }
  +\cos\theta}
\over {2\sqrt{ \cos^2\theta + P^2 \sin^2\theta }}}
\qquad\mbox{and}\qquad
\sin^2\gamma = {{ \sqrt{ \cos^2\theta + P^2
\sin^2\theta }
  -\cos\theta}
\over {2\sqrt{ \cos^2\theta + P^2 \sin^2\theta }}}\;.
$$
In this, the 2$\times$2 matrix representation here is based on the
choice
\be
\ket+= \pmatrix{ 1\cr 0} \qquad\mbox{and} \qquad \ket-= \pmatrix{
0\cr1}\;.
\ee
For $\Delta$, we have
\be
\Delta=-{1\over2} \pmatrix{ P(1+\cos\theta) & \sin\theta \cr
\sin\theta & P(1-\cos\theta)}\;.
\ee
Then by calculating $U^\dagger\Delta U$ and substituting
\be
\langle\omega_i\vert\Delta\vert\omega_j\rangle =\langle i\vert
U^\dagger\Delta U\vert j\rangle
\ee
into Eq.~(\ref{super-op_rep}), we get
\be
X = \pmatrix{ X_{++} & X_{+-}\cr X_{-+}
& X_{--}}\;,
\ee
where
\be
X_{\pm\pm}=\mp {{P\cos\theta\big(1-(1-P^2)\sin^2\theta\big)}\over
{\cos^2\theta + P^2 \sin^2\theta}}
\ee
and
\be
X_{+-}=X_{-+}={{-\sin\theta (\cos^2\theta + P^2
\sin^2\theta)}\over {\cos^2\theta + P^2 \sin^2\theta}}
\ee
This can be diagonalized by
\be
U_0= \pmatrix{ \cos\gamma_0 & \sin\gamma_0 \cr -\sin\gamma_0
& \cos\gamma_0}
\ee
with
$$
\cos^2\gamma_0 = { { \sqrt{ P^2 \cos^2\theta + \sin^2\theta }
  -P\cos\theta}
\over {2\sqrt{ P^2 \cos^2\theta + \sin^2\theta }} }
\qquad\mbox{and}\qquad
\sin^2\gamma_0 = { { \sqrt{ P^2 \cos^2\theta + \sin^2\theta }
  +P\cos\theta}
\over {2\sqrt{ P^2 \cos^2\theta + \sin^2\theta }} }
$$
such that
\be
 X =
\sum_{i=+,-}\nu_i\proj{\nu_i}, \quad \ket{\nu_\pm}=U_0\ket{\pm},
\ee
where $\nu_\pm= \sqrt{ P^2 \cos^2\theta + \sin^2\theta }$.
Therefore the optimal POVM constitutes a set of the two rank-one
elements $E_1=\proj{\nu_+}$ and $E_2=\proj{\nu_-}$.

For such $\{E_b\}$,
\be
\langle\nu_\pm\vert \rho \vert\nu_\pm\rangle = { {\sqrt{ P^2
\cos^2\theta
         + \sin^2\theta }\mp P}
\over {2\sqrt{ P^2 \cos^2\theta + \sin^2\theta }} },
\ee
and hence we have
\be
\left[
  {{\tr(E_b\Delta)}\over{\tr(E_b\rho)}}
\right]^2 =P^2 \cos^2\theta + \sin^2\theta \quad{\mbox{for }} b=1,
2.
\ee
Thus the equality condition Eq.~(\ref{equality_condition1}) is
also satisfied.

Putting all this together, we obtain that the accessible fidelity
for $\cal P$ is given by
\be
F_{\cal P}={1\over2}\!\left( 1+ \sqrt{ \cos^2\theta + (P^2 {\rm
cos}^2\theta + \sin^2\theta) \sin^2\theta } \right).
\ee
The quantumness of the set ${\cal
S}=\{\ket{\psi_1},\ket{\psi_2}\}$ is now the minimum of $F_{\cal
P}$ with respect to the parameter $P$. In this case, the minimum
is clearly attained for $\pi_1=\pi_1=1/2$, and so
\bea
Q({\cal S})&=&\min_{\{\pi_i\}}F_{\cal P}
\nonumber\\
&=&
{1\over2}\!\left( 1+ \sqrt{ \cos^2\theta + \sin^4\theta } \right).
\eea
In terms of $x=|\langle\psi_0|\psi_1\rangle|$, we get the
expression assumed without proof in Ref.~\cite{Fuchs00}.  Namely,
\be
Q({\cal S}) = \frac{1}{2}\Big(1+\sqrt{1-x^2+x^4}\Big)\;.
\label{BigJohn}
\ee

Notice that if we consider varying the parameter $x$ in
Eq.~(\ref{BigJohn}), two states will be the most quantum with
respect to each other when $x=1/\sqrt{2}\,$. In that case,
$Q\approx0.933$. Therefore, in opposition to clonability, this
measure does have the pleasing feature that two states are the
most quantum with respect to each other when $\theta=45^\circ$.
Thus $Q({\cal S})$ fulfills the slogan set forth in Section 1.

It is worth noting that the optimal measurement for the accessible
fidelity is also the one that maximizes the average success
probability in hypothesis testing, Eq.~(\ref{Kurt}),
\be
P_s=\pi_1 + \tr(E_0 \Delta)\;.
\ee
In fact,
we can see that
\be
\Delta=\omega_{0+}\proj{\nu_+}
          +\omega_{0-}\proj{\nu_-},
\ee
where
\be
\omega_{0\pm}={1\over2} \left( -P \pm\sqrt{ P^2 \cos^2\theta +
\sin^2\theta } \right).
\ee
The projectors $\{\proj{\nu_+}, \proj{\nu_-}\}$ define the
Helstrom measurement~\cite{Helstrom76}.

\subsection{Accessible Fidelity for Group Symmetric Cases of a Qubit}

Deriving the quantumness for sets $\cal S$ consisting of more then
three elements remains an open problem in general.  However, the {\it
accessible fidelity\/} can be found for certain sets with symmetry
properties. In this section, we consider group symmetric sources on a
qubit and derive optimal strategies for achieving the accessible
fidelity in these cases.  What are particularly interesting are the
subcases where an arbitrary, simple von Neumann measurement will do.
This suggests a connection between these cases and the Scrooge
ensembles considered in Ref.~\cite{Jozsa94b}.

\subsubsection{Real symmetric states}

Let us first consider the $M$-ary real symmetric source on a
qubit. Introducing the rotation operator
\begin{equation}
V\equiv{\rm exp}(-i{\pi\over M} \sigma_y) = \pmatrix{ \cos
{{\pi}\over M} & -\sin {{\pi}\over M}\cr \sin {{\pi}\over M} &
\cos {{\pi}\over M} },
\end{equation}
we make the ensemble consisting of the $M$ states
\begin{equation}
\label{psik} \ket{\psi_k} =   V^k \ket{\psi_0}
= \pmatrix{ \cos \frac{k\pi}{M}\cr \sin \frac{k\pi}{M} }, \quad
k=0, \ldots , M-1,
\end{equation}
taken with equal prior probabilities $\pi_i = \frac{1}{M}$. On the
Bloch sphere, these states $\Pi_k=\proj{\psi_k}$ are equally
spaced around a great circle in the $x$-$z$ plane consisting of
all real states.

The accessible fidelity for this set was already obtained by using
a simple differentiation technique based on Lagrange multipliers
in Ref.~\cite{Barnett01}. Here, we derive the optimal solution by
a different method. Namely, we use the symmetry of the source,
along with the generalization of Davies' theorem proved in Section
3.

By that theorem, the optimal POVM for accessible fidelity can be
specified by a 3-output rank-1 POVM. We parameterize such POVMs
${\cal E}=\{E_b=\proj{E_b}\}$ by~\cite{Sasaki99}
\begin{eqnarray}
\vert E_0\rangle&=& \sqrt{2-\alpha^2 -\beta^2 }
\pmatrix{\cos\theta
\cr
              \sin\theta},
 \\
\vert E_1\rangle&=&\alpha
  \pmatrix{  \cos(\theta+\varphi_\alpha) \cr
             \sin(\theta+\varphi_\alpha)},
 \\
\vert E_2\rangle&=&\beta
  \pmatrix{  \cos(\theta+\varphi_\beta) \cr
             \sin(\theta+\varphi_\beta)},
\end{eqnarray}
where the coefficients are given by
\be
\alpha^2={{\cos\varphi_\beta}\over
            {\sin\varphi_\alpha
             \sin(\varphi_\alpha-\varphi_\beta)}}
\qquad\mbox{and}\qquad
\beta^2={{\cos\varphi_\alpha}\over
            {\sin\varphi_\beta
             \sin(\varphi_\beta-\varphi_\alpha)}},
\ee
and
\begin{equation}
0\le\alpha^2+\beta^2\le2.
\end{equation}
It follows then that
\begin{equation}\label{POVM_Eb}
  E_b=\frac{\lambda_b}2
\left(  I+ \sigma_x\sin\Theta_b
            + \sigma_z\cos\Theta_b\right),
\end{equation}
where
\begin{equation}
\lambda_b= \left \{
\begin{array}{ll}
2-\alpha^2-\beta^2& \quad (b=0) \cr \alpha^2& \quad (b=1) \cr
\beta^2& \quad (b=2)
\end{array}\right]
\qquad\mbox{and}\qquad
\Theta_b= \left \{
\begin{array}{ll}
2\theta& \quad (b=0) \\
2\theta+2\varphi_\alpha& \quad (b=1) \\
2\theta+2\varphi_\beta& \quad (b=2)
\end{array}\right].
\end{equation}
Evaluating the mapping operators
\be
  M_b=\sum_i \pi_i \tr(\Pi_i E_b)\Pi_i
\ee
for obtaining the achievable fidelity
\be
F({\cal E})=\sum_b\, \lambda_1\!\!\left(  M_b\right)\;,
\label{William2}
\ee
we see that they can be written as
\begin{equation}
  M_b=\frac{\lambda_b}{4}
\left(  I+ \sigma_x{1\over2}\sin\Theta_b
     + \sigma_z{1\over2}\cos\Theta_b\right).
\end{equation}
Note in particular that these operators have no dependence at all
on $M$.

Thus, for $M\ge3$, the $M_b$ can all be diagonalized by the same
operators,
\begin{equation}\label{zm}
P_b =\left(
\begin{array}{cc}
\cos {{\Theta_b}\over2} & -\sin {{\Theta_b}\over2}  \\
\sin {{\Theta_b}\over2} & \cos {{\Theta_b}\over2}
\end{array}
\right),
\end{equation}
as
\begin{equation}
P_b^\dagger M_b P_b =\left(
\begin{array}{cc}
\frac{3\lambda_b}{8} & 0  \\
0 & \frac{\lambda_b}{8}
\end{array}
\right).
\end{equation}

Hence it follows that the accessible fidelity is
\begin{equation}\label{Facc_max_M=3}
F=\max_{\{E_b\}} \sum_b{3\over8}\lambda_b={3\over4}.
\end{equation}
This value can be achieved by {\it any\/} 3-outcome POVM with rank-1
elements!  In particular, the degenerate case where two of the
three outcomes are identical---i.e., the measurement is a standard
von Neumann measurement---will also do.

If the prior probabilities of the signals $\pi_i$ were not equal,
things would not have proceeded in such a straightforward way.

\subsubsection{Platonic Solids}

The second example for which we can derive an exact expression for
the accessible fidelity has to do with sets of group symmetric
states characterized by groups with an irreducible unitary
representation. For such ensembles, Davies considered the
maximization problem of the Shannon mutual information and derived
the existence of a class of POVMs with the same symmetry. The
mutual information, like the achievable fidelity, is also a convex
function of POVMs. Therefore the spirit of Davies' work can be
applied to accessible fidelity as well. In fact we can derive a
similar result about the possible forms of the optimal POVM by
slight modifications of the proof given in Ref.~\cite{Davies78}.

Let $\{U_g \;\vert\; g\in G\}$ be a projective unitary
representation of the group $G$ characterizing the ensemble $\cal
P$ of signal states $\{\Pi_i=\proj{\psi_i}\}$ with equal prior
probabilities $\pi_i=1/M$. Suppose a POVM $\{
A_a=\kappa_a\proj{v_a}\}$ $(\vert\langle v_a\vert
v_a\rangle\vert=1)$ and the assignment
$\{a\mapsto\ket{\varphi_a}\}$ are optimal, that is
\begin{equation}
F_{\cal P} ={1\over M}\sum_a\sum_i\tr \left(  A_a  \Pi_i\right)
\vert\langle\varphi_a\vert\psi_i\rangle\vert^2.
\end{equation}
Consider another POVM and assignment
\begin{eqnarray}
  B_{ag}
&=&{1\over{\vert G\vert}}  U_g   A_a   U_g^\dagger,
\label{MilesDavis1}\\
ag
&\mapsto&
  U_g\ket{\varphi_a},
\label{MilesDavis2}
\end{eqnarray}
where $\vert G\vert$ is the number of group elements. Look at the
quantity
\begin{equation}
F(\{  B_{ag}\}) =\sum_g{1\over M}\sum_a\sum_i \tr\left(  B_{ag}
\Pi_i\right) \vert\langle\varphi_a\vert
U_g^\dagger\vert\psi_i\rangle\vert^2.
\end{equation}
This reduces to
\begin{eqnarray}
F(\{  B_{ag}\})
&=&{1\over{\vert G\vert}}
   \sum_g{1\over M}\sum_a
   \langle\varphi_a\vert
   \sum_i   U_g^\dagger \Pi_i  U_g
            A_a
            U_g^\dagger \Pi_i  U_g
   \vert\varphi_a\rangle
\nonumber\\
&=&{1\over{\vert G\vert}}
   \sum_g{1\over M}\sum_a
   \langle\varphi_a\vert
   \sum_{i'}  \Pi_{i'}
            A_a
           \Pi_{i'}
   \vert\varphi_a\rangle
\nonumber\\
&=&{1\over{\vert G\vert}}
   \sum_g F_{\cal P}
\nonumber\\
&=&F_{\cal P}.
\end{eqnarray}
Thus the strategy of Eqs.~(\ref{MilesDavis1}) and
(\ref{MilesDavis2}) is optimal too. Let us next define
\begin{equation}
  C_{g}^{(a)}\equiv{d\over{\vert G\vert}}
  U_g \proj{v_a}   U_g^\dagger,
\label{POVM_symmetric}
\end{equation}
where $d$ is the dimension of the Hilbert space spanned by the
$\ket{\psi_i}$. Due to the irreducibility of $\{  U_g \;\vert\;
g\in G\}$,
\begin{equation}
\sum_g  C_{g}^{(a)}=  I,
\end{equation}
holds for each $a$, that is, $\{  C_{g}^{(a)}\}$ is a POVM for
each $a$. $\{  B_{ag}\}$ can be written as
\begin{equation}
\sum_{ag}  B_{ag} =\sum_{ag}{\kappa_a\over d}  C_{g}^{(a)} =  I,
\end{equation}
that is $\{  B_{ag}\}$ is a convex combination of $\{
C_{g}^{(a)}\}$. Then by convexity
\begin{equation}
F(\{  B_{ag}\}) \le \sum_{ag}{\kappa_a\over d}F(\{
C_{g}^{(a)}\}).
\end{equation}
Since $\{  B_{ag}\}$ is optimal, there must exist an optimal POVM
of the form Eq.~(\ref{POVM_symmetric}). This means that there
exists a rank-1 nucleus from which the optimal POVM can be
generated by applying the group elements $U_g$.

Let us apply this result to qubit sources corresponding to the
regular polyhedra, i.e., the five Platonic solids. These sources
include the tetrahedron ($\vert G\vert=4$), octahedron ($\vert
G\vert=6$), cube ($\vert G\vert=8$), icosahedron ($\vert G\vert=12$),
and dodecahedron ($\vert G\vert=20$).  These are characterized by the
regular polyhedral groups in three dimensional Euclidean space
corresponding to their Bloch sphere representation.

Let the signal states be
\be
\Pi_g={1\over2}\left(I+\vec\psi_g\cdot\vec\sigma\right), \quad
\vec\psi_g=(\alpha_g, \beta_g, \gamma_g)^T,
\ee
From the above result, there must exist the optimal POVM for the
accessible fidelity of the form
\begin{equation}\label{POVM_covariant}
  E_g={2\over{\vert G\vert}}U_g\proj{e}U_g^\dagger.
\end{equation}
Obviously the mapping operator is also group covariant, and the
accessible fidelity can simply be given by
\be
F=\vert G\vert\lambda_1(M_0),
\ee
where $g=0$ is an element of the group. Denoting
\be
\proj{e}={1\over2}\left(I+\vec e\cdot\vec\sigma\right), \quad \vec
e=(x,y,z)^T,
\ee
the mapping operator can be obtained as
\be
M_0=\frac{1}{2\vert G\vert^2} \left[ \vert G\vert I + \sum_g
(\vec\psi_g\cdot\vec e)(\vec\psi_g\cdot\vec\sigma) \right],
\ee
where $\sum_g\vec\psi_g=0$ has been used. Defining the 3-by-3
matrix
\be
\Psi\equiv \sum_g \vec\psi_g \vec\psi_g^T,
\ee
this can be further modified as
\be
M_0=\frac{1}{2\vert G\vert^2} \left[ \vert G\vert I + (\Psi \vec
e)\cdot\vec\sigma \right].
\ee
From the irreducibility of group representation in the three
dimensional Euclidean space, we can see (Shur's lemma)
\be
\Psi=\frac{\vert G\vert}{3} \left(
\begin{array}{ccc}
1 & 0 & 0 \\
0 & 1 & 0 \\
0 & 0 & 1 \\
\end{array}
\right).
\ee
Then
\be
M_0=\frac{1}{2\vert G\vert} \left[
 I
+ {1\over3}\vec e\cdot\vec\sigma \right].
\ee
The maximum eigenvalue and its eigenstate are $2/3\vert G\vert$ and
$\ket e$, respectively. Therefore the optimal strategy consists of
the measurement $\{E_g\}$ described by Eq. (\ref{POVM_covariant}) for
Eves and the state assignment $g\mapsto U_g\ket e$ for Yves. The
choice of $\ket e$ can be arbitrary. The accessible fidelity is
$F=2/3$ for all the Platonic solids. We can also see by simple
calculation that the strategy consisting of any von Neumann
measurement for Eve and the output of the corresponding eigenstate by
Yves, realizes the same limit as well.

Again, the calculations in this section do not solve the problem of
the quantumness for the Platonic solids.  For that we would have to
first find the accessible fidelity for these states with arbitrary
probabilities $\pi_i$.

\subsection{Accessible Fidelity for Unitarily Invariant Ensembles}

The most interesting result of the last subsection---i.e., that the
accessible fidelity for the Platonic solids is insensitive to which
measurement is used by Eve---is reminiscent of an observation due to
Barnum in Ref.~\cite{Barnum02} connecting information-disturbance
problems for the uniform ensemble with those for ``spherical
2-designs in $CP_{d-1}$ where $d$ is a power of an odd prime.'' Thus,
let us focus on the case where $\cal P$ consists of a continuous
infinity of states distributed according to the unitarily invariant,
or ``uniform,'' distribution on ${\cal H}_d$.  To promote further
work on Barnum's idea and to motivate a conjecture in the next
section, let us here tabulate the accessible
fidelity
for the
ensemble $\cal P$ in the alternative cases of real Hilbert-space,
complex Hilbert-space, and quaternionic Hilbert-space quantum
mechanics~\cite{Araki80,Stueckelberg60,Adler95}. The complex case, of
course, has been studied before in great detail by various methods
(see Ref.~\cite{Bruss99} and references therein). Here, we extend,
the method of Barnum in Ref.~\cite{Barnum98} to the real and
quaternionic cases as well.

All of these cases can be subsumed into a single formalism by
denoting states in the following way:
\be
\ket{\psi}=\sum_{j=1}^d\!\left(\sum_{k=1}^\nu x_{\! jk}\,\vec{
e}_k\right)\!\ket{\phi_j}
\ee
where $\vec{e}_1=\vec{1}$, $\vec{e}_2=\vec{i}$, $\vec{e}_3=\vec{j}$,
and $\vec{e}_4=\vec{k}$ are the four quaternionic basis states, and
$\ket{\phi_j}$ is a fixed orthonormal basis on ${\cal H}_d$. In real
Hilbert-space quantum mechanics $\nu=1$, in complex $\nu=2$, and in
quaternionic $\nu=4$. The unitarily invariant measure $d\Omega_\psi$
on the pure states in ${\cal H}_d$ is then given by~\cite{Jones94},
\be
d\Omega_\psi=\pi^{-\nu d/2}\,\Gamma\!\left(\frac{\nu d}{2}\right)
\delta\!\left(1-\sum_{j=1}^d\sum_{k=1}^\nu x_{\! jk}^2\right)\!
\prod_{j=1}^d\prod_{k=1}^\nu dx_{\! jk}\;,
\ee
where $\Gamma(x)$ is the usual gamma function.

To set up the problem, it turns out to be less convenient in this
case to work with the prepackaged formula Eq.~(\ref{William}) than to
go back to the basic expression in Eq.~(\ref{Jens}).  Thus consider
Eve performing a general rank-one POVM $E_b = g_b |b\rangle\langle
b|$ and Yves generating the states $\sigma_b =
|\phi_b\rangle\langle\phi_b|$ in response.  Since $\sum_b E_b = I$,
it follows that $\sum_b g_b =d$. The fidelity under this strategy is
then given by
\be
F_{\cal P}({\cal E},{\cal M})=\sum_b g_b \int |\langle
b|\psi\rangle|^2 |\langle\phi_b|\psi\rangle|^2 d\Omega_\psi\;.
\ee
This integral can be evaluated by the methods of Jones~\cite{Jones94}
since, as he has shown for any sufficiently smooth function $f$, the
integral $J$ becomes
\bea
J &=& \int |\langle\phi|\omega\rangle|^2
f\big(|\langle\psi|\omega\rangle|^2\big)d\Omega_\omega
\\
&=&
\frac{1}{d-1}\left[\big(1-|\langle\psi|\phi\rangle|^2\big) \int
f\big(|\langle\psi|\omega\rangle|^2\big)d\Omega_\omega
\nonumber
+ \big(d|\langle\psi|\phi\rangle|^2-1\big) \int
|\langle\psi|\omega\rangle|^2
f\big(|\langle\psi|\omega\rangle|^2\big)d\Omega_\omega\right]
\eea
Now using the formula~\cite{Jones94}
\be
\int |\langle\phi|\psi\rangle|^{2n} d\Omega_\psi= \frac{%
\Gamma\Big(\frac{\nu d}{2}\Big)\Gamma\Big(\frac{\nu}{2}+n\Big)}{%
\Gamma\Big(\frac{\nu}{2}\Big)\Gamma\Big(\frac{\nu d}{2}+n\Big)}\;,
\ee
we can see after a little algebraic rearrangement that this quantity
is maximized by taking $|\phi_b\rangle=|b\rangle$. Therefore
\be
F_{\cal P}= \frac{
d\, \Gamma\Big(\frac{\nu d}{2}\Big)\Gamma\Big(\frac{\nu}{2}+2\Big)}{%
\Gamma\Big(\frac{\nu}{2}\Big)\Gamma\Big(\frac{\nu d}{2}+2\Big)}\;,
\ee
whose values reduce to
\be
\frac{3}{d+2},\qquad\frac{2}{d+1},\qquad\mbox{and}\qquad
\frac{3}{2d+1}
\label{Mucus}
\ee
in the real, complex and quaternionic cases respectively.

Note how the accessible fidelities for the real symmetric states and
for the Platonic solids in the last two subsections did indeed match
the values $3/(d+2)$ and $2/(d+1)$.

\subsection{Lower Bound to Accessible Fidelity for the Lifted Trine States}

As a final example, let us consider the ensemble of lifted trine
states in ${\cal H}_3$, each given with probability $1/3$. This
ensemble is of particular interest because of the discovery by
Shor~\cite{Shor00,Shor02} that its accessible information requires a
measurement consisting of six outcomes, even though the number of
inputs is only three. That is, in this case, the number of outputs
must be greater than the number of inputs to get the most
information!

The lifted trine states are defined by
\begin{eqnarray}
|\psi_0\rangle &=& \Big(\sqrt{1-\alpha}, 0, \sqrt{\alpha}\Big)
\nonumber
\\
|\psi_1\rangle &=& \Big(-{\textstyle{\frac{1}{2}}}\sqrt{1-\alpha},
{\textstyle{\frac{\sqrt{3}}{2}}}\sqrt{1-\alpha}, \sqrt{\alpha}\Big)
\\
|\psi_2\rangle &=& \Big(-{\textstyle{\frac{1}{2}}}\sqrt{1-\alpha}, -
{\textstyle{\frac{\sqrt{3}}{2}}}\sqrt{1-\alpha}, \sqrt{\alpha}\Big)
\nonumber
\end{eqnarray}
where the ``lifting parameter'' $\alpha$ can range between 0 and 1.
Shor has shown that when $\alpha<0.061$, there is a unique POVM
optimal for achieving the accessible information of this ensemble.
The POVM ${\cal E}_{\rm\scriptscriptstyle
SHOR}=\{E_b=|e_b\rangle\langle e_b|\}$ consists of six elements of
the form
\bea
|e_1\rangle &=& c \, (0,1,0) \nonumber\\
|e_2\rangle &=& c\!\left(\textstyle{\frac{\sqrt{3}}{2}},
\textstyle{\frac{1}{2}},0\right) \nonumber\\
|e_3\rangle &=& c\!\left(\textstyle{\frac{\sqrt{3}}{2}},
\textstyle{-\frac{1}{2}},0\right) \nonumber\\
&&\\
|e_4\rangle &=& d \, (1,0,x) \nonumber\\
|e_5\rangle &=& d \!\left(\textstyle{-\frac{1}{2}},
\textstyle{\frac{\sqrt{3}}{2}},x\right) \nonumber\\
|e_6\rangle &=& d \!\left(\textstyle{-\frac{1}{2}},
\textstyle{\frac{\sqrt{3}}{2}},x\right)
\nonumber
\eea
where the normalization constants
\be
c^2=\frac{2}{3}\left(1-\frac{1}{2x^2}\right)\qquad\mbox{and}\qquad
d^2=\frac{1}{3x^2}
\ee
are dependent upon a parameter $x$, yet to be fixed.  A numerical
optimization gives that
\be
x=\tan \phi_\alpha
\ee
and
\be
\sin^2\phi_\alpha \approx \frac{1-\alpha}{1+29.591\alpha}\;.
\ee
The discrepancy between the mutual information given by this POVM and
the mutual information obtained from the optimal von Neumann
measurement occurs very near $\alpha=1/40$.  Therefore let us
consider the achievable fidelity that comes about due to this
measurement at $\alpha=1/40$.  With a little numerical work, one can
extract that $F({\cal E}_{\rm\scriptscriptstyle SHOR})\approx
0.79999$.

On the other hand, it turns out to be possible to find an analytic
expression for the achievable fidelity derived from the square-root
measurement for all values of $\alpha$.  Using Eq.~(\ref{Jeroen}),
one finds
\bea
F_{\scriptscriptstyle\rm SRM}
&=&
\frac{1}{8}\left\{\rule{0mm}{5mm}
3+\alpha^2+2\sqrt{2}\alpha^{1/2}(1-\alpha)^{3/2} + \right.
\label{Maudlin}
\\
&&
\left.\left[9-24\alpha+126\alpha^2-200\alpha^3+105\alpha^4 +
4\sqrt{2}\alpha^{1/2}(1-\alpha)^{3/2}(3+8\alpha-15\alpha^2)\right]^{1/2}\,\right\}.
\nonumber
\eea
At $\alpha=1/40$ this evaluates to $F_{\scriptscriptstyle\rm
SRM}\approx 0.84766$.  Thus the square-root measurement is a better
measurement for achievable fidelity than the measurement optimal for
mutual information.  This contrasts with all the previous examples,
where the measurement optimal for mutual information was {\it also\/}
optimal for accessible fidelity.

Of course, this does not answer what the accessible fidelity actually
is for the lifted trines.  It is interesting to explore
however---under the {\it supposition\/} that the square-root
measurement might be optimal and that quantumness might be achieved
with equal prior probabilities $\pi_i$---for what value of $\alpha$
the trine states are the most quantum with respect to each other.
Minimizing Eq.~(\ref{Maudlin}) over all $\alpha$, we get:
\be
\min_\alpha F_{\scriptscriptstyle\rm SRM} =
\frac{3}{4}\qquad\mbox{when}\qquad\alpha=0\;,
\ee
which simply signifies that the states are most quantum with respect
to each other when they are no longer linearly independent.   There
is another interesting regime, however, between $\alpha=1/3$ and
$\alpha=1$---that is, when the lifted trines are restricted to having
angles between $0^\circ$ and $90^\circ$ from each other. Then,
\be
\min_{\frac{1}{3}\le\alpha\le1} F_{\scriptscriptstyle\rm SRM} \approx
0.89682\qquad\mbox{when}\qquad\alpha\approx 0.78868\;.
\ee
For this value of $\alpha$ the trine states are all $46.92^\circ$
from each other.  Thus we note a possibly interesting trend:  By
using larger dimensional Hilbert spaces, we may be able to achieve
greater levels of quantumness even while using closer-to-orthogonal
states.

\section{Conclusions and Open Questions} \label{TheEnd}

A quick summary of the results shown in this paper can be found in
the following table:

\begin{center}
\begin{tabular}{|c||c|} \hline
& \\
formula for achieveable fidelity &
Eq.~(16) \\
& \\
\hline
& \\
optimal measurements exist / Davies-like theorem
& Section 3 \\
& \\
\hline
& \\
lower bounds on accessible fidelity
& Eqs.~(27), (29) and (31) \\
& \\
\hline
& \\
formula for accessible fidelity and quantumness & \\
of two nonorthogonal states
&
Eqs.~(71) and (73) \\
& \\
\hline
& \\
accessible fidelity for M-ary real symmetric source & \\
on a qubit & Eq.~(91) \\
& \\
\hline
& \\
accessible fidelity for Platonic solids on a qubit &
below Eq.~(109) \\
& \\
\hline
& \\
accessible fidelity for unitarily invariant ensembles & \\
(real, complex, and quaternionic) &
Eq.~(116) \\
& \\
\hline
& \\
lower bound to accessible fidelity of lifted-trine states &
Eq.~(122) \\
& \\
\hline
\end{tabular}
\end{center}
\bigskip

What are the practical uses for a measure of quantumness?  At least
two stand out.  As mentioned in the Introduction, it has become
common in experimental verifications of quantum teleportation to
tabulate an input-output fidelity for defining a threshold over which
teleportation can be said to have been achieved.  (See
Ref.~\cite{Braunstein00} for an extended discussion.)  The reasoning
behind this is the following.

Perfection in the laboratory is in principle unattainable.  So, in a
proposed experimental demonstration of quantum teleportation, when
can on say that something nonclassical has been achieved? There are
many markers of ever more exacting requirement, but perhaps one of
the most basic is that quantum entanglement has actually been used by
the procedure.  How could one tell if it had or had not been? For
this, the fidelity between input and output is an appropriate tool.
If Alice and Bob were to use nothing more than a classical
communication channel between them, the best teleportation fidelity
they could achieve would be given by the Eve-Yves scenario described
in the Introduction.  Here is where our work has some impact.

For any ensemble $\cal E$ to be teleported, the minimum
fidelity-threshold required to certify that entanglement has actually
been used is $F({\cal E})$. The payoff of using a maximally quantum
set of states in an actual laboratory procedure is that it will
define the minimum threshold---a threshold that is presumably the
easiest to achieve.  The difference is stark: If the set of states an
experimentalist tests his teleportation device upon consists solely
of two nonorthogonal states, then he will need to achieve a fidelity
of at least 0.933 to be able to declare success. On the other hand,
if he were to use a unitarily invariant ensemble, then he would only
need to achieve a fidelity of $2/(d+1)\le0.666$.  The payoff becomes
bigger and bigger with higher dimension.  If one could find a finite
set of states that achieves the same fidelity threshold, that would
be even better. (See Question 2 below.)

In some cases though---for instance in continuous variable quantum
teleportation~\cite{Furusawa98}---it may be difficult for the
experimenter to prepare arbitrary ensembles.  In that case it is nice
to have a tool for evaluating the quality of one ensemble over
another.  For example, it is known that a ``uniform'' ensemble of
coherent states gives a threshold of $1/2$~\cite{Braunstein00}.  But
what might the ability to introduce a small amount of squeezing do
for the threshold?  These are the sort of questions that can be
explored.

Beyond teleportation, the most important use of quantumness may be in
the line of theoretical investigations of quantum key distribution.
Of course, quantum cryptography is an already very developed field,
and it would be a bit brash to assume that this somewhat ad hoc
quantity may have any direct connection to the subject.  There are
already several established security quantities actively explored in
the field. (See, for instance, Ref.~\cite{GilbertBig}.) Nevertheless,
studying quantumness may give insight and quick information about the
desirability of pursuing more exotic protocols based on more exotic
quantum-state alphabets than have been pursued so
far.\footnote{Though work in that direction is starting to
accumulate.  See, Refs.~\cite{Bourennane02,Bruss02}.}

In particular, one might consider the tradeoff between the security
achievable by an alphabet and the rate at which legitimate users can
accumulate key with it.  For instance, consider a B92 protocol using
an alphabet of two quantum states with overlap $x$.  The quantumness
of this alphabet, as we have already shown, is
\be
Q(x) = \frac{1}{2}\Big(1+\sqrt{1-x^2+x^4}\Big)\;.
\ee
On the other hand, the probability that Bob will be able to
unambiguously identify the signal Alice sent him (i.e., by using an
optimal unambiguous state discrimination scheme \cite{Peres93}) will
be
\be
P(x) = 1-x
\ee
Thus, though two states may be the ``most secure'' against
eavesdropping when $x=1/\sqrt{2}$, they will certainly not be optimal
for sending a raw bit of key in the absence of eavesdropping.

A ``quick and dirty'' figure of merit for the tradeoff between the
two desirable features is the product of the security and the rate:
\be
T(x)=P(x)[1-Q(x)]\;.
\ee
In other words, this is the product of Bob's probability of
reproducing Alice's bit and Eve's probability of being caught---we
would like to simultaneously maximize the two if we could, but we
cannot. One can check numerically that the figure of merit is
maximized when
\be
x\approx 0.54807\qquad\mbox{i.e.,}\qquad \phi\approx56.77^\circ\;.
\ee
The value $T(x)$ gives at that point is $0.02514$.

Now compare this to what we would find {\it if\/} Eq.~(\ref{Maudlin})
happened to denote the actual quantumness $Q(\alpha)$ of the lifted
trine states. The optimal unambiguous state discrimination
probability $P(\alpha)$ for that ensemble is
\cite{Chefles98,Obajtek01}
\be
P(\alpha)= \left \{
\begin{array}{ll}
3\alpha & \quad\mbox{if}\quad 0\le\alpha\le\frac{1}{3}
\cr
\frac{3}{2}(1-\alpha) & \quad\mbox{if}\quad \frac{1}{3}\le\alpha\le 1
\rule{0mm}{5mm}
\end{array}\right].
\ee
To put the problem on the same footing as the B92 protocol, we can
take the figure of merit $T(\alpha)=P(\alpha)[1-Q(\alpha)]$. This
quantity is optimized at the value $T(\alpha)\approx 0.04105$---over
1.6 times the two-state value---when $\alpha\approx 0.68535$.  In
terms of the angle between the three vectors, the best figure of
merit is reached when $\phi=58.13^\circ$.  Thus we see an improvement
in the figure of merit, at the same time as the signal states are
more orthogonal to each other.

This seems significant, and one has to wonder whether the trend might
continue with higher dimensions.  That is, whether one might be able
to use ever more orthogonal signals and still be able to achieve
better and better figures of merit in the security/bit-rate tradeoff.
Also one might wonder whether a similar effect could carry over to an
ensemble of $N$ symmetric coherent states, as in
Ref.~\cite{vanEnk02}.  That would be very nice because one could
contemplate B92-like protocols for coherent states with very high
photon number.  But this is sheer speculation at this point.

In any case, as it ought to be clear from our exposition, the
properties of accessible fidelity and quantumness are far from being
exhaustively explored. Here is a list of open questions that strike
us as the most important presently.

\begin{enumerate}

\item  Though we defined the notion of the quantumness $Q_d$ of a Hilbert
space ${\cal H}_d$ in Eq.~(\ref{Charles}), we did not in the end make
any use of it. A natural guess for the value is that it is simply
equal to the accessible fidelity for the unitarily invariant
ensemble:
\be
Q_d\stackrel{?}{=}\frac{2}{d+1}\;.
\ee
Is this so?  Similarly, one can ask the same question for
real-Hilbert space quantum mechanics and quaternionic quantum
mechanics, based on the expressions in Eq.~(\ref{Mucus}).  Sections
(4.2.1) and (4.2.2) provide some warrant for this speculation.

\item
If it is so, can this value always be achieved by a finite ensemble
of signalling states?  What is the minimum number of such states?  Is
it $d^2$?  There is some evidence that this might be the case from
the following.

Suppose there exists a symmetric informationally complete
POVM.\footnote{We stress, however, that this assumption is not
trivial.  Several colleagues, including R.~Blume-Kohout, C.~M. Caves,
G.~G. Plunk, M.~Fickus, J.~Renes, A.~J. Scott, and W.~K. Wootters,
have spent some time toward trying to give an existence proof for
these objects---only so far to meet defeat. On the other hand,
explicit constructions have been found in $d=2,3,4$ and numerical
examples have been found for dimensions up to
$d=14$~\cite{Blume-Kohout03}. To our knowledge, the concept was first
introduced by C.~M. Caves, Ref.~\cite{CavesSICPOVM}.}\ \ \ That is,
there exists $d^2$ vectors $|\psi_i\rangle$ on ${\cal H}_d$ such that
the operators
\be
E_i = \frac{1}{d}|\psi_i\rangle\langle\psi_i|
\ee
form a POVM and
\be
{\rm tr}(E_i E_j)=\frac{1}{d^2(d+1)}\quad\forall\; i\ne j\;.
\ee
If we consider a scenario in which Alice's signal ensemble consists
of the states $|\psi_i\rangle$, $\pi_i=1/d^2$, and the corresponding
operators $E_i$ specify Eve's measurement, the achievable fidelity
given by Eq.~(\ref{William}) evaluates to
\bea
F({\cal E})
&=&
\sum_j\,\lambda_1\!\!\left(\sum_i\frac{1}{d^2}{\rm tr}(\Pi_i E_j)
\Pi_i\right)
\nonumber\\
&=&
\sum_j\,\lambda_1\!\!\left(\sum_i{\rm tr}(E_i E_j) E_i\right)
\nonumber\\
&=&
\sum_j\,\lambda_1\!\!\left(\frac{1}{d^2}E_j+\frac{1}{d^2(d+1)}\sum_{i\ne
j} E_i\right)
\nonumber\\
&=&
\sum_j\,\lambda_1\!\left(\frac{1}{d^2}E_j+\frac{1}{d^2(d+1)}(I-E_j)\right)
\nonumber\\
&=&
\lambda_1\!\left(\frac{1}{d+1}I+\frac{d}{d+1}E_k\right)\qquad\mbox{for
any } E_k
\nonumber\\
&=&
\frac{2}{d+1}\;.
\eea

Of course, this proves nothing---no optimizations have been carried
out, either over the measurements or over the probability
distributions $\{\pi_i\}$.  Furthermore, it is not known if $Q_d$ is
bounded below by $2/(d+1)$.  The ensemble above is offered solely as
a candidate for further thought.

\item
Are there any useful upper and lower bounds to accessible fidelity
$F_{\cal P}$?

\item
Are there any ensembles $\cal P$ for which an optimal
hypothesis-testing measurement is {\it not\/} also optimal for
achieving the accessible fidelity?

\item
In analogy to Shor's demonstration of the peculiar properties of
accessible information for the lifted trine states, are there any
ensembles $\cal P$ on ${\cal H}_d$ consisting of $d$ elements but for
which a POVM of $d^2$ elements is required to achieve its accessible
fidelity?

\item
Is accessible fidelity multiplicative?  That is, for two ensembles
${\cal P}=\{\pi_i,\Pi_i\}$ and ${\cal S}=\{\phi_j,\Phi_j\}$ on ${\cal
H}_1$ and ${\cal H}_2$, respectively, let us define an ensemble
${\cal P}\otimes{\cal S}$ on ${\cal H}_1\otimes{\cal H}_2$ by
\be
{\cal P}\otimes{\cal S}\equiv\{\pi_i\phi_j,\Pi_i\otimes\Phi_j\}\;.
\ee
The question is,
\be
F_{{\cal P}\otimes{\cal S}}=F_{\cal P}F_{\cal S}\;?
\ee
It is not particularly difficult to show that accessible information
is additive for independent ensembles~\cite{Holevo73a}.  But it seems
that a wholly new technique may be needed for the present question.
If it does turn out that accessible fidelity is not multiplicative,
then this might be pause to reconsider whether the present definition
of quantumness is the most appropriate one.

\item
For the purposes of quantum cryptography, ensembles $\cal
P_{\rm\scriptscriptstyle LI}$ consisting of linearly independent
states $|\psi_i\rangle$ take on a special role.  This is because the
states in those ensembles can be unambiguously identified with some
finite probability. In particular, such ensembles can serve as the
basis for various generalized B92-like protocols.  Therefore it seems
of interest to define the LI-quantumness for a Hilbert space in
analogy to the overall quantumness
\be
Q_{\rm\scriptscriptstyle LI}(d)=\inf_{{\cal P}_{\rm\scriptscriptstyle
LI}} F_{{\cal P}_{\rm\scriptscriptstyle LI}}\;.
\ee
How does $Q_{\rm\scriptscriptstyle LI}(d)$ compare to $Q_d$?  Is
$Q_{\rm\scriptscriptstyle LI}(d)$ just the $Q_{d^\prime}$ of a
smaller dimensional space?
\end{enumerate}

Finally, one can ponder the deeper significance of the underlying
fact in quantum mechanics that quantumness attempts to quantify:  the
quantum world is a world ``sensitive to the touch''~\cite{Fuchs02}.
What this might mean for all of us, the field of quantum information
is only starting to tell.

\section{Acknowledgments}
Much of this work was carried out at the California Institute of
Technology in Spring 1999 and at Tamagawa University in Spring 2000
and Spring 2001.  To this end, CAF acknowledges the support of a Lee
DuBridge Fellowship at Caltech, and the warm hospitality of the
Research Center for Quantum Communication at Tamagawa University
during those periods.  He also thanks R.~Obajtek for many discussions
during her Bell Labs summer internship, 2001.  MS thanks S.~M.
Barnett and C.~Gilson for valuable discussions.  We both thank an
anonymous referee for many helpful suggestions concerning the
presentation.


\begin{thebibliography}{99}

\bibitem{Glauber63}
R.~J. Glauber, ``The Quantum Theory of Optic Coherence,'' Phys.\
Rev.\ {\bf 130}, 2529--2539 (1963); R.~J. Glauber, ``Coherent and
Incoherent States of the Radiation Field,'' Phys.\ Rev.\ {\bf
131}, 2766--2788 (1963).

\bibitem{Wootters82}
W.~K. Wootters and W.~H. Zurek, ``A Single Quantum Cannot Be
Cloned,'' Nature {\bf 299}, 802--803 (1982).

\bibitem{Dieks82}
D.~Dieks, ``Communication by EPR Devices,'' Phys.\ Lett.\ A {\bf
92}, 271--272 (1982).

\bibitem{Yuen86}
H.~P. Yuen, ``Amplification of Quantum States and Noiseless Photon
Amplifiers,'' Phys.\ Lett.\ A {\bf 113}, 405--407 (1986).

\bibitem{Fuchs00}
C.~A. Fuchs, ``Just {\it Two\/} Nonorthogonal Quantum States'' in
{\sl Quantum Communication, Computing, and Measurement 2}, edited by
P.~Kumar, G.~M. D'Ariano, and O.~Hirota (Kluwer, Dordrecht, 2000),
pp.~11--16.

\bibitem{Buzek96}
V.~Bu\v{z}ek and M.~Hillery, ``Quantum Copying:\ Beyond the
No-Cloning Theorem,'' Phys.\ Rev.\ A {\bf 54}, 1844--1852 (1996).

\bibitem{Bruss98}
D.~Bru\ss, D.~P. DiVincenzo, A.~Ekert, C.~A. Fuchs,
C.~Macchiavello, and J.~A. Smolin, ``Optimal Universal and
State-Dependent Quantum Cloning,'' Phys.\ Rev.\ A {\bf 57},
2368--2378 (1998).

\bibitem{Holevo73a}
A.~S. Holevo, ``Information-Theoretical Aspects of Quantum
Measurement,'' Prob.\ Inf.\ Trans.\ {\bf 9}, 110--118 (1973).

\bibitem{Helstrom76}
C.~W. Helstrom, {\sl Quantum Detection and Estimation Theory},
(Academic Press, NY, 1976).

\bibitem{Fuchs96a}
C.~A. Fuchs, {\sl Distinguishability and Accessible Information in
Quantum Theory}, Ph.D. thesis, University of New Mexico, 1996. LANL
archive {\tt quant-ph/9601020}.

\bibitem{Ekert94}
A.~K. Ekert, B.~Huttner, G.~M. Palma, and A.~Peres,
``Eavesdropping on Quantum Cryptographical Systems,'' Phys.\ Rev.\
A {\bf 50}, 1047--1056 (1994).

\bibitem{Schumacher95}
B.~Schumacher, ``Quantum Coding,'' Phys.\ Rev.\ A {\bf 51},
2738--2747 (1995).

\bibitem{Massar95}
S. Massar and S. Popescu, ``Optimal Extraction of Information from
Finite Quantum Ensembles,'' Phys.\ Rev.\ Lett.\ {\bf 74}, 1259--1263
(1995).

\bibitem{Derka98}
R. Derka, V. Bu\v{z}ek, and A. K. Ekert, ``Universal Algorithm for
Optimal Estimation of Quantum States from Finite Ensembles via
Realizable Generalized Measurement,'' Phys.\ Rev.\ Lett.\ {\bf 80},
1571--1575 (1998).

\bibitem{Latorre98}
J. I. Latorre, P. Pascual, and R. Tarrach, ``Minimal Optimal
Generalized Quantum Measurements,'' Phys.\ Rev.\ Lett.\ {\bf 81},
1351--1354 (1998).

\bibitem{Bruss99}
D.~Bru\ss\ and C.~Macchiavello, ``Optimal State Estimation for
d-Dimensional Quantum Systems,'' Phys.\ Lett.\ A {\bf 253}, 249--251
(1999).

\bibitem {Vidal99}
G. Vidal, J.~I. Latorre, P. Pascual, and R. Tarrach, ``Optimal
Minimal Measurements of Mixed States,'' Phys.\ Rev.\ A {\bf 60},
126--135 (1999).

\bibitem {Acin99}
A. Ac{\'\i}n, J. I. Latorre, and P. Pascual, ``Optimal Generalized
Quantum Measurements for Arbitrary Spin Systems,'' Phys.\ Rev.\ A
{\bf 61}, 022113/1--7 (2000).

\bibitem {Sasaki01PhasEst}
M. Sasaki, A. Carlini, and A. Chefles, ``Optimal Phase Estimation and
Square Root Measurement,'' J. Phys.\ A {\bf 34}, 7017--7027 (2001).

\bibitem{Chefles00}
A.~Chefles, ``Quantum State Discrimination,'' Contemp.\ Phys.\
{\bf 41}, 401--424 (2000).

\bibitem{Barnum00}
H.~Barnum, P.~Hayden, R.~Jozsa, and A.~Winter, ``On the Reversible
Extraction of Classical Information from a Quantum Source,'' Proc.\
R.\ Soc.\ Lond.\ A {\bf 457}, 2019--2039 (2001).

\bibitem{Hayden02}
P.~Hayden, R.~Jozsa, and A.~Winter, ``Trading Quantum for Classical
Resources in Quantum Data Compression,'' J.\ Math.\ Phys.\ {\bf 43},
4404--4444 (2002).

\bibitem{Boschi98}
D.~Boschi, S.~Branca, F.~De~Martini, L.~Hardy, and S.~Popescu,
``Experimental Realization of Teleporting an Unknown Pure Quantum
State via Dual Classical and Einstein-Podolsky-Rosen Channels,''
Phys.\ Rev.\ Lett.\ {\bf 80}, 1121--1125 (1998).

\bibitem{Furusawa98}
A.~Furusawa, J.~L. S{\o}rensen, S.~L. Braunstein, C.~A. Fuchs,
H.~J. Kimble, and E.~S. Polzik, ``Unconditional Quantum
Teleportation,'' Science {\bf 282}, 706--709 (1998).

\bibitem{Braunstein00}
S.~L. Braunstein, C.~A. Fuchs, and H.~J. Kimble, ``Criteria for
Continuous-Variable Quantum Teleportation,'' J. Mod.\ Opt.\ {\bf 47},
267--278 (2000).

\bibitem{Bennett93}
C.~H. Bennett, G.~Brassard, C.~Cr\'{e}peau, R.~Jozsa, A.~Peres,
and W.~K. Wootters, ``Teleporting an Unknown Quantum State via
Dual Classical and Einstein-Podolsky-Rosen Channels,'' Phys.\
Rev.\ Lett.\ {\bf 70}, 1895--1899 (1993).

\bibitem{Jozsa94}
R.~Jozsa, ``Fidelity for Mixed Quantum States,'' J.\ Mod.\ Opt.\ {\bf
41}, 2315--2323 (1994).

\bibitem{Schumacher90}
B.~Schumacher, ``Information from Quantum Measurements,'' in {\sl
Complexity, Entropy and the Physics of Information}, edited by
W.~H. Zurek (Addison-Wesley, Redwood City, CA, 1990), pp.~29--37.

\bibitem{Bhatia97}
R.~Bhatia, {\sl Matrix Analysis}, (Springer, New York, 1997).

\bibitem{Holevo73b}
A.~S. Holevo, ``Statistical Decision Theory for Quantum Systems,''
J.\ Mult.\ Anal.\ {\bf 3}, 337--394 (1973).

\bibitem{Fujiwara98}
A.~Fujiwara and H.~Nagaoka, IEEE Trans.\ Inf.\ Theory {\bf 44},
1071--1086 (1998).

\bibitem{Davies78}
E.~B. Davies, ``Information and Quantum Measurement,'' IEEE
Trans.\ Inf.\ Theory {\bf IT-24}, 596--599.

\bibitem{Sasaki99}
M. Sasaki, S. M. Barnett, R. Jozsa, M. Osaki, and O. Hirota,
``Accessible Information and Optimal Strategies for Real Symmetric
Quantum Sources,'' Phys.\ Rev.\ A {\bf 59}, 3325--3335 (1999).

\bibitem{Shor00}
P.~W. Shor, ``On the Number of Elements Needed in a POVM Attaining
the Accessible Information,'' to appear in {\sl Quantum
Communication, Computing, and Measurement 3}, edited by O. Hirota and
P.~Tombesi (Kluwer, Dordrecht, 2001).  See also {\tt
quant-ph/0009077}.

\bibitem{Renyi66}
A.~R\'enyi, ``On the Amount of Missing Information and the
Neyman-Pearson Lemma,'' in {\sl Research Papers in Statistics:\
Fest\-schrift for J. Neyman}, edited by F.~N. David (Wiley, New
York, 1966), pp.~281--288.

\bibitem{Holevo78}
A.~S. Holevo, ``On Asymptotically Optimal Hypothesis Testing in
Quantum Statistics,'' Theor.\ Prob.\ Appl.\ {\bf 23}, 429--432
(1978).

\bibitem{Hausladen94}
P.~Hausladen and W.~K. Wootters, ``A `Pretty Good' Measurement for
Distinguishing Quantum States,'' J. Mod.\ Opt.\ {\bf 41},
2385--2390 (1994).

\bibitem{BraunsteinCaves94}
S.~L. Braunstein and C.~M. Caves, ``Statistical Distance and the
Geometry of Quantum States,'' Phys.\ Rev.\ Lett.\ {\bf 72},
3439--3443 (1994).

\bibitem{FuchsCaves94}
C.~A. Fuchs and C.~M. Caves, ``Ensemble-Dependent Bounds for
Accessible Information in Quantum Mechanics,'' Phys.\ Rev.\ Lett.\
{\bf 73}, 3047--3050 (1994).

\bibitem{Jozsa94b}
R.~Jozsa, D.~Robb, and W.~K. Wootters, ``Lower Bound for Accessible
Information in Quantum Mechanics,'' Phys.\ Rev.\ A {\bf 49}, 668--677
(1994).

\bibitem{Barnett01}
S. M. Barnett, C. R. Gilson and M. Sasaki, ``Fidelity and the
Communication of Quantum Information,'' J. Phys.\ A {\bf 34},
6755--6766 (2001).

\bibitem{Barnum02}
H.~Barnum, ``Information-Disturbance Tradeoff in Quantum Measurement
on the Uniform Ensemble and on the Mutually Unbiased Bases,'' {\tt
quant-ph/0205155}.

\bibitem{Araki80}
H. Araki, ``On a Characterization of the State Space of Quantum
Mechanics,'' Comm.\ Math.\ Phys.\ {\bf 75}, 1 (1980).

\bibitem{Stueckelberg60}
E.~C.~G. Stueckelberg, ``Quantum Theory in Real Hilbert Space,''
Helv.\ Phys.\ Acta {\bf 33}, 727 (1960).

\bibitem{Adler95}
S.~L. Adler, {\sl Quaternionic Quantum Mechanics and Qua\-ntum
Fields}, (Oxford U. Press, New York, 1995).

\bibitem{Barnum98}
H.~N. Barnum, {\sl Quantum Information Theory}, Ph.~D. thesis,
University of New Mexico, 1998.

\bibitem{Jones94}
K.~R.~W. Jones, ``Fundamental Limits Upon the Measurement of State
Vectors,'' Phys.\ Rev.\ A {\bf 50}, 3682--3699 (1994); K.~R.~W.
Jones, ``Entropy of Random Quantum States,'' J.\ Phys.\ A {\bf 23},
L1247--L1251 (1990); K.~R.~W. Jones, ``Quantum Limits to Information
about States for Finite Dimensional Hilbert Space,'' J.\ Phys.\ A
{\bf 24}, 121--130 (1991).

\bibitem{Shor02}
P.~W. Shor, ``The Adaptive Classical Capacity of a Quantum Channel,
or Information Capacities of Three Symmetric Pure States in Three
Dimensions,'' quant-ph/0206058.

\bibitem{GilbertBig}
G.~Gilbert and M.~Hamrick, ``Practical Quantum Cryptography: A
Comprehensive Analysis (Part One),'' {\tt quant-ph/0009027}.

\bibitem{Bourennane02}
M.~Bourennane, A.~Karlsson, G.~Bj\"ork, N.~Gisin, and N.~J. Cerf,
``Quantum Key Distribution using Multilevel Encoding: Security
Analysis,'' J.\ Phys.\ A {\bf 35}, 10065--10076 (2002);  N.~J. Cerf,
M.~Bourennane, A.~Karlsson, and N.~Gisin, ``Security of Quantum Key
Distribution Using d-level Systems,'' Phys.\ Rev.\ Lett.\ {\bf 88},
127902/1--4 (2002).

\bibitem{Bruss02}
D.~Bru\ss\ and C.~Macchiavello, ``Optimal Eavesdropping in
Cryptography with Three-dimensional Quantum States,''  Phys.\ Rev.\
Lett.\ {\bf 88}, 127901/1--4 (2002).

\bibitem{Peres93}
A.~Peres, {\sl Quantum Theory: Concepts and Methods}, (Kluwer,
Dordrecht, 1993).

\bibitem{Chefles98}
A.~Chefles and S.~M. Barnett, ``Optimum Unambiguous State
Discrimination Between Linearly Independent Symmetric States,''
Phys.\ Lett.\ A {\bf 250}, 223--229 (1998).

\bibitem{Obajtek01}
R.~A. Obajtek, ``A Few Aspects of Three-State Quantum Cryptography,''
summer intern report, Bell Labs, 2001.

\bibitem{vanEnk02}
S. J. van Enk, ``Unambiguous State Discrimination of Coherent States
with Linear Optics: Application to Quantum Cryptography,'' Phys.\
Rev.\ A {\bf 66}, 042313/1--5 (2002).

\bibitem{Blume-Kohout03}
R. Blume-Kohout and J. Renes, private communication (Spring, 2003).

\bibitem{CavesSICPOVM}
C.~M. Caves, ``Symmetric Informationally Complete POVMs,'' UNM
Information Physics Group internal report, posted at {\tt
http://info.phys.unm.edu/$\,\tilde{\;\;}\!$caves/reports/reports.html},
(1999).

\bibitem{Fuchs02}
C.~A. Fuchs, ``Quantum Mechanics as Quantum Information (and only a
little more),'' {\tt quant-ph/ 0205039}.

\end{thebibliography}
\end{document}